\documentclass[unnumsec,webpdf,contemporary,large]{oup-authoring-template}%

\graphicspath{{Figures/}}

 \usepackage{booktabs}
 \usepackage{makecell}

\theoremstyle{thmstyleone}%
\theoremstyle{thmstyletwo}%
\theoremstyle{thmstylethree}%

\begin{document}
\onecolumn

\journaltitle{Integrative and Comparative Biology}
\DOI{DOI added during production}
\copyrightyear{2026}
\pubyear{2026}
\vol{XX}
\issue{X}
\access{Advance Access publication date added during production}
\appnotes{Original Article}

\firstpage{1}


\title[Buoyancy-dependent flow generation]{Buoyancy-dependent flow generation by collectively migrating swimmers}

\author[1]{Nina Mohebbi}
\author[1,2,$\ast$]{John O. Dabiri}

\address[1]{\orgdiv{Graduate Aerospace Laboratories}, \orgname{California Institute of Technology}, \orgaddress{Pasadena, CA 91125, USA}}

\address[2]{\orgdiv{Mechanical and Civil Engineering}, \orgname{California Institute of Technology}, \orgaddress{Pasadena, CA 91125, USA}}

\corresp[$\ast$]{Corresponding author. \href{mailto:nina_mohebbi@brown.edu}{nina\_mohebbi@brown.edu}}



\abstract{Collective vertical swimming may generate aggregate-scale flows that contribute to mixing and transport in stratified environments. The strength of these flows depends not only on swimmer behavior but also on environmental properties. Here we examine how fluid density affects flow generated by vertically migrating swarms of brine shrimp \textit{Artemia salina}. Using simultaneous three-dimensional swimmer tracking and particle image velocimetry, we measured swimmer kinematics and the induced flow field during phototactically driven migrations under four controlled salinity conditions.
Induced velocity increased with buoyancy forcing and scaled with the parameter $N(\rho_s - \rho)$, where $N$ is the number of swimmers and $\rho_s - \rho$ is the density difference between swimmers and the surrounding fluid ($R^2 = 0.70$, $p = 5.9 \times 10^{-5}$). A multiple regression including swimmer number, swimming speed, fluid density, and the swarm Gaussian width confirmed that density remained a significant predictor of induced velocity after controlling for the other variables ($p = 0.012$, $R^2 = 0.82$). A simplified actuator-disk model captures the first-order dependence of induced velocity on buoyancy forcing and swimmer momentum, suggesting that environmentally driven density variations can substantially modify the hydrodynamic impact of collective vertical migration.} 


\maketitle

\section{Introduction}

Diel Vertical Migration (DVM) is a daily synchronized movement of aquatic animals which occurs daily across most of the global ocean, particularly in regions of strong stratification and active mixing, such as continental shelf breaks, pycnocline zones, and mesopelagic layers~\citep{bandara_two_2021}. By biomass, DVM is the largest migration that has been measured on earth~\citep{bandara_two_2021} and the vertical distances traveled during a DVM event span hundreds of meters. These distances often coincide with permanent pycnoclines, which range in depth from approximately 200 to 800 meters, and separate lighter surface waters from denser deep waters with density changes between 1 and 2 percent~\citep{feucher_subtropical_2019,sverdrup_oceans_1942}. Therefore, understanding the relationship between swimmer buoyancy and induced hydrodynamics may be essential to predict the impact that vertical migrations have on ocean mixing and nutrient transport. 

Self-propelled neutrally buoyant organisms with no net acceleration impart no net momentum to the surrounding fluid because thrust balances drag~\citep{lighthill_mathematical_1997}. Such momentum-less wakes decay within a few body lengths, and any mixing relies on secondary processes such as viscosity-enhanced drift volumes or turbulence~\citep{katija_viscosity-enhanced_2009, visser_biomixing_2007}. However, if a swimmer is denser than the ambient fluid, an additional upward thrust component is required to offset its net-negative buoyancy. That surplus thrust drives a downward jet with finite momentum, transforming the wake from momentum-less to momentum-bearing.

Many marine organisms possess tissues that are denser than water and have various adaptations to maintain near neutral buoyancy during vertical movement in the water. These adaptations include swim bladders in teleost fish, oil-rich livers in sharks, and gas-filled chambers in cephalopods~\citep{alexander_buoyancy_1982}. Active changes in an organism's buoyancy can be used to adjust vertical position without swimming, as in the case of a nonmotile phytoplankton, \textit{Pyrocystis noctiluca}, which has been found to use rapid cell inflation to enable long-distance vertical migration without the use of swimming appendages~\citep{larson_inflation-induced_2024}. The majority of the biomass in DVM are planktonic animals~\citep{katija_biogenic_2012}, many of which may depend solely on swimming to maintain their vertical position in the water~\citep{alexander_buoyancy_1982}. For example, krill species are negatively buoyant due to limited lipid reserves and a dense skeletal structure, and use continuous active swimming and intermittent sinking behavior to maintain or adjust their vertical position~\citep{tarling_oceanic_2017}. 

In this paper, induced vertical migrations of brine shrimp, \textit{Artemia salina}, were carried out while varying the salinity of the surrounding water to modulate the buoyant force on the swimmer. Although \textit{A. salina} do not perfectly replicate oceanic zooplankton communities, brine shrimp swim within the same laminar to transitional Reynolds regime as typical oceanic zooplankton such as copepods and krill which migrate at intermediate Reynolds numbers ranging from $\sim$1 to 1000~\citep{yen_life_2000,catton_quantitative_2007,murphy_hydrodynamics_2013}. In addition, brine shrimp provide a tractable experimental system for studying collective swimming flows because they exhibit strong positive phototaxis, enabling repeatable induced migrations under laboratory conditions and controlled investigation of how density differences between swimmers and surrounding fluid influence aggregation-scale flows. 

Simultaneous 3-D swimmer tracking with 2-D 2-component flow measurements across a range of salinity conditions was used to empirically quantify the impact on induced collective flow during vertical migration. This approach treats salinity as a controlled mechanism for varying buoyancy under laboratory conditions, with the understanding that salinity-correlated behavioral changes are possible. Therefore, we measured swimmer number, swimming speed, and swarm distribution across conditions and used a regression-based analysis to isolate the buoyancy contribution from these correlated factors.  We frame our findings accordingly: as evidence for buoyancy-modulated flow generation in a controlled system rather than as an exhaustive separation of all salinity-related effects. Our results suggest that increasing swimmer buoyancy reduces aggregate-induced flow velocity. To contextualize and extend these results to oceanic conditions, the results are applied to a theoretical model that estimates the aggregation as a simplified disk-shaped source of momentum. According to this model, changes in buoyancy predictably change the strength of the collective induced flow, with implications spanning a wide range of ocean water densities.

\section{Materials and Methods} \label{sec:methods}

\subsection{Experimental Setup} \label{sec:method_tank}
All experiments were conducted in a transparent acrylic tank measuring 1.2~m in height with a square cross-section of 0.5~m $\times$ 0.5 m. The tank was filled with artificial seawater prepared from Instant Ocean Sea Salt (Spectrum Brands) dissolved in deionized (DI) water; depending on the experiment, the working salinity ranged from 15~parts~per~thousand~(ppt) to 22~ppt. The water was kept at approximately 20$^{\circ}$C (68$^{\circ}$F), and two sides of the tank as well as the tank lid were insulated with 12~mm closed-cell polyethylene foam (McMaster-Carr 9349K5, 1.25~cm thick, thermal conductivity 0.27~W~m$^{-1}$~K$^{-1}$). The tank lid also limited evaporation from the surface. These steps were taken to avoid the convection flows that are triggered from vertical temperature gradients in this geometry. For flow visualization, the tank was seeded with 10 $\mu$m silver coated glass microspheres (Potters CONDUCT-O-FIL). 

\subsection{Induced migration protocol}
Adult \textit{Artemia salina} (1~cm body length (BL)) were obtained from Carolina Biological Supply (Burlington, NC, USA) and all experiments were conducted within 24 hours after animals were received. The animals were added to the tank in densely packed 0.25 tsp (1.2 mL, $\approx$125 individuals) increments. 

\begin{figure}
 \centering{\includegraphics[scale=0.24]{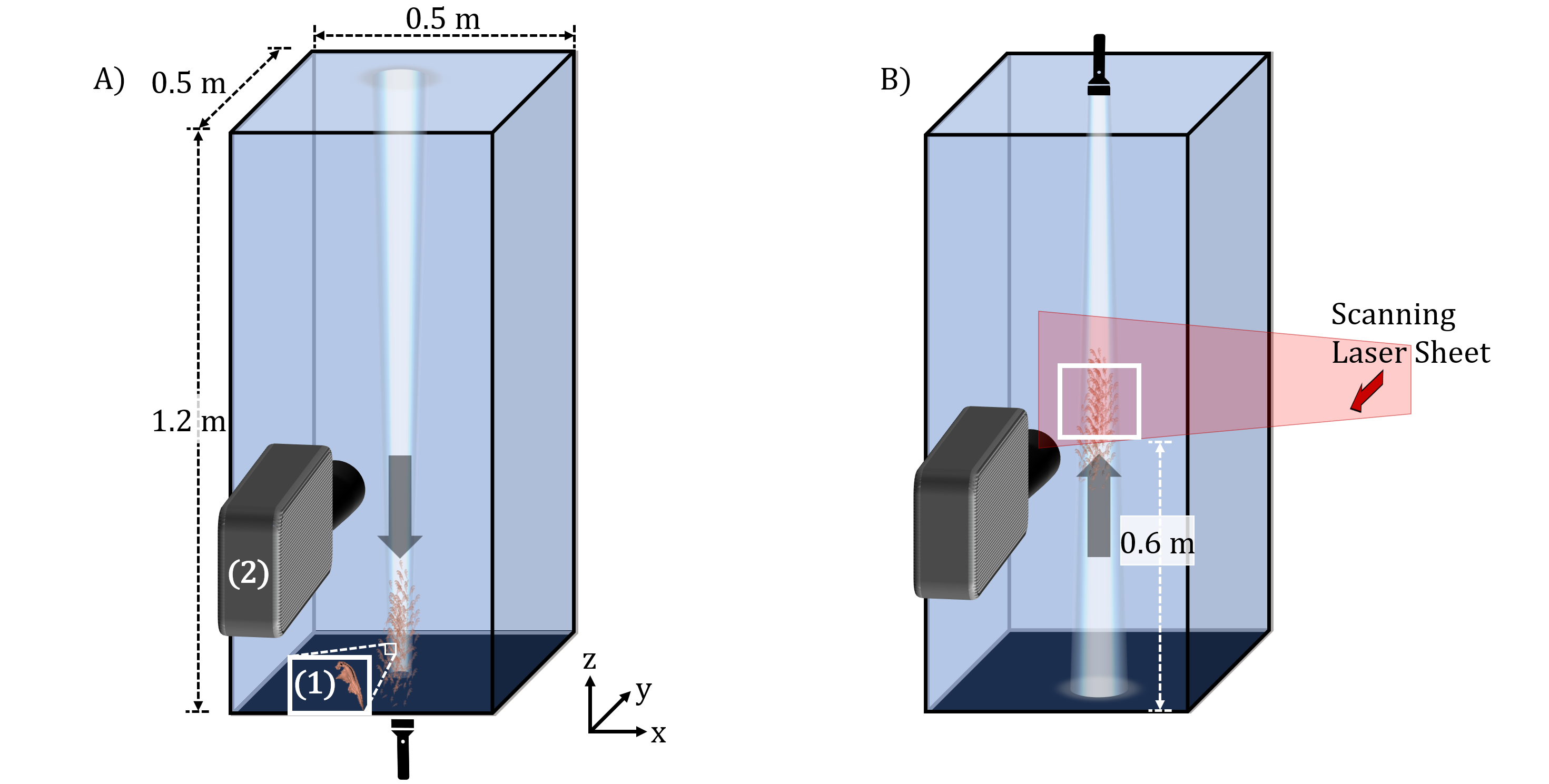}}
     \caption{Experimental setup. (A) Schematic of the test tank (0.5~m $\times$ 0.5~m $\times$ 1.2 m) used to induce upward collective vertical migration in \textit{Artemia salina} (1). A flashlight (bottom) provides a positive phototactic cue, eliciting downward swimming and aggregation initialized at the tank floor. A high-speed camera (2) images the central scanning volume. (B) After allowing the tank to settle for a minimum of 15 minutes, quiescence is confirmed and a vertical migration is induced by switching the target light to one on top of the tank. As the swimmers migrate up they pass through the imaging region. A scanning laser sheet illuminates the swimmers and tracer particles moving along the $y$-axis. The white box indicates the region 0.6~m above the tank floor that is imaged. Coordinate axes ($x, y, z$) are oriented as shown, with $z$ aligned with gravity.}
\label{fig:method_exp}
\end{figure}

Brine shrimp exhibit positive phototaxis, meaning that they swim towards sources of light. An LED flashlight placed below the tank was used to attract the brine shrimp to the tank floor in preparation for an induced migration. Then, the fluid in the tank was allowed to settle for a minimum of 15 minutes (figure~\ref{fig:method_exp}A). A 25~s video of the laser-sheet-illuminated tank was recorded every 5 min and analyzed using PIVlab in MATLAB~\citep{thielicke_pivlab_2014} until the maximum time-averaged vertical velocity was less than 0.02~cm~s$^{-1}$. Once this criterion for quiescence was met, an upward migration was initiated. To induce a migration, the bottom flashlight was turned off, and simultaneously, a flashlight (PeakPlus LFX1000) mounted above the water tank was turned on, causing the brine shrimp to swim up (figure~\ref{fig:method_exp}B).

The geometry of the swarm, such as aspect ratio, distance traveled, and packing fraction, differs from those observed in DVM. The aspect ratio (height/width) of oceanic zooplankton aggregations typically ranges from about 0.03 to 0.2. Field acoustics show that large-area aggregations form flat layers, whereas smaller compact swarms can be longer relative to their width~\citep{nicol_living_2003,hamner_behavior_1983}. In our laboratory setting, swimmer distributions resembled the smaller, longer, and compact swarms. The swarm spanned the length of the tank vertically, resulting in an aspect ratio of $\sim$5 using the height of the tank and width of scanning window. Packing fractions in the ocean rarely exceed 1\%, typically ranging from 0.001\% to around 0.1\%. In the experiments packing fraction is $\sim$1\%. Lastly, the ratio of vertical migration distance to the swimmer body length typically falls between $10^{3}$ and $10^{5}$~\citep{hays_review_2003,ringelberg_diel_2009}, in our experiments the swimmers travel 120 body lengths. 

\subsection{Imaging methods and analysis} \label{sec:3D_method} 
To capture three-dimensional swimming trajectories during vertical migration, a 3-D imaging system~\citep{fu_single-camera_2021} was implemented using high-speed scanning with a planar laser combined with a single high-speed camera (Photron FASTCAM SA-Z). A continuous-wave red laser (671 nm, 5 W) sheet was vertically translated through the depth of the tank by rotating a mirror mounted on a galvanometric scanner (Thorlabs GVS211/M) driven by an externally controlled voltage signal. 

The scanning system was calibrated using a custom-built three-dimensional marker array consisting of 1.6~mm opaque spheres embedded in a transparent 6 × 6 × 6 grid, forming a regular lattice within an 8~cm acrylic cube. This calibration object is reconstructed in 3-D and used to define the mapping between image coordinates and physical space. This procedure also corrected for optical distortions introduced by the imaging path. The system was previously validated by comparing experimental measurements of a laminar free jet at intermediate Reynolds number (Re $\approx$ 50) against direct numerical simulations of synthetic tracer fields, demonstrating the method's accuracy for both position and velocity reconstruction~\citep{fu_single-camera_2021}.

\begin{figure}
 \centering{\includegraphics[scale=0.24]{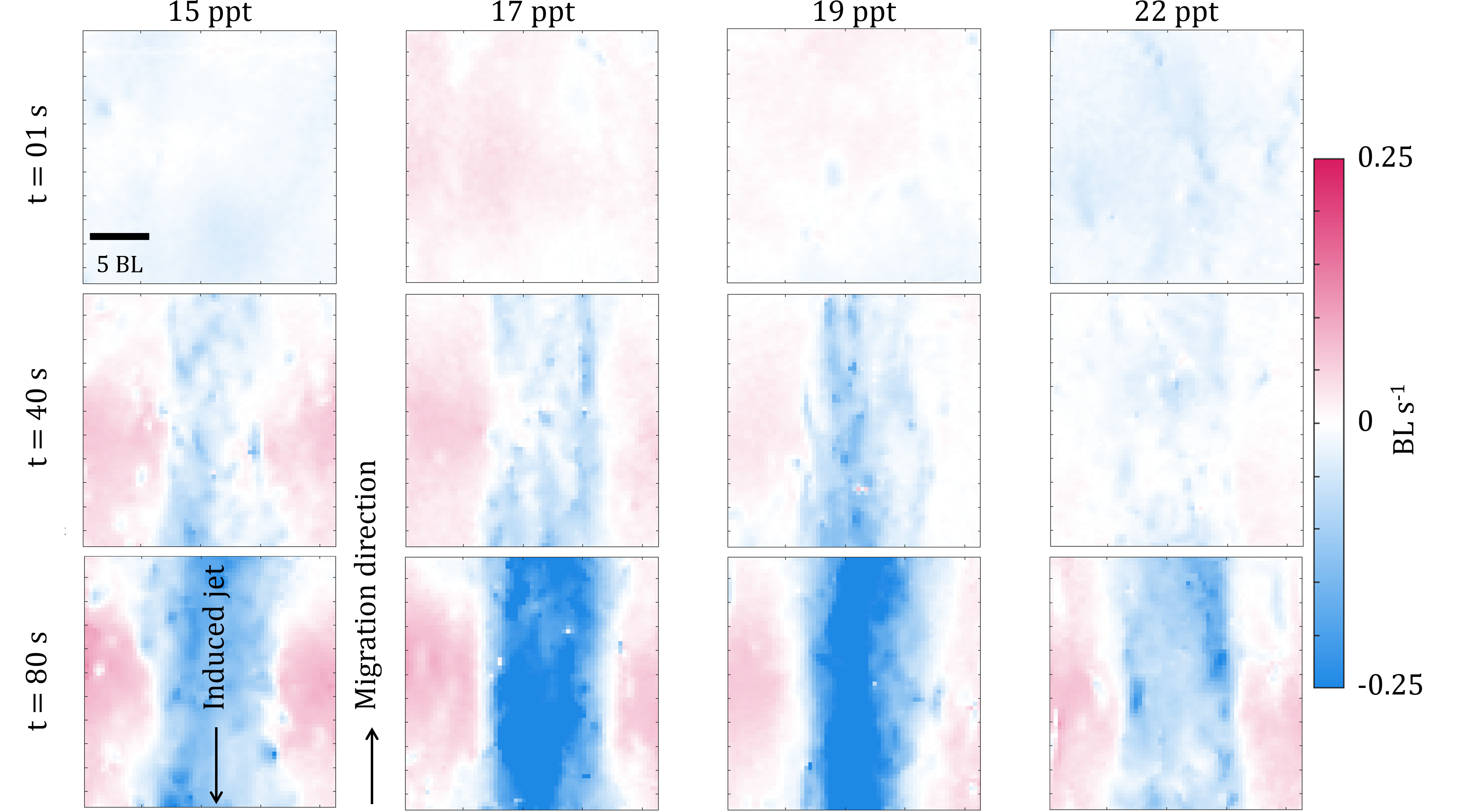}}
     \caption{Representative PIV vertical-velocity fields during induced migrations at four salinity conditions, 15, 17, 19, and 22~ppt. Images are arranged in columns corresponding to each salinity condition and rows showing snapshots at times t=1, 40, 80 s. Swimmer body length (BL) is 1~cm. Arrows are included to indicate flow direction.}
\label{fig:method_2DPIV}
\end{figure}

Given that swimmer ascent speeds were approximately 1~cm~s$^{-1}$, displacement during a single scan remained under 0.2~cm in all cases. This is small relative to the animal body length, 1~cm, so each scan was treated as an instantaneous volume. Four vertical migration trials were performed per salinity condition. For each experimental condition, 3-D data were collected from a fixed imaging region positioned near the tank center with a width ($x$) and height ($z$) of 22~BL, and depth ($y$) of 2.5~BL. Representative PIV vertical-velocity fields obtained at the four salinity conditions are shown in figure~\ref{fig:method_2DPIV}.

\subsection{Salinity treatments and density measurement} \label{sec:method_salinity}
Four salinities were chosen to test swimmer buoyancy as a variable: 15, 17, 19, and 22~parts~per~thousand~(ppt). These salinities correspond to densities of approximately 1009, 1011, 1012.5, and 1014.5~kg~m$^{-3}$, respectively. The densities selected are based on previous tests in which a visible jet was produced, since all conditions had fluid density lower than that of the swimmers. \textit{A. salina} typically live in environments of 35~ppt, if we assume they are neutrally buoyant in this environment we can estimate their density as 1025~kg~m$^{-3}$; the surrounding fluid in our experimental conditions are thus 1 to 1.5$\%$ less dense than the swimmers across our experimental range. This range of experimental conditions also brackets the typical density variation that organisms may encounter during diel vertical migrations in the ocean, a 1\% change in fluid density ~\citep{bandara_two_2021,sverdrup_oceans_1942}.

Salinity levels were adjusted by dissolving different quantities of Instant Ocean into DI water (Type II, ChemWorld) and measured using a CastAway Conductivity, Temperature and Depth Profiler (CTD) (salinity range = 0-42~ppt, precision = 0.10~ppt, resolution = 0.01~ppt, temperature range = -5-45 \textdegree C, accuracy = 0.05°C, resolution = 0.01°C, www.sontek.com). For each test, water of the desired salinity was prepared the night before brine shrimp were received. The following morning, measurements were taken with the CTD at 5 locations in the tank (4 corners and center), and then the animals were loaded into the tank. The densities for the four conditions, averaged across depth and tank locations, were: 1009.3 $\pm$.02, 1010.9 $\pm$.04, 1012.7 $\pm$.02, and 1014.4 $\pm$.03~kg~m$^{-3}$. For readability, these four conditions will be referred to as 15, 17, 19, and 22~ppt.

\subsection{Data Processing}
\begin{figure}
 \centering{\includegraphics[scale=0.225]{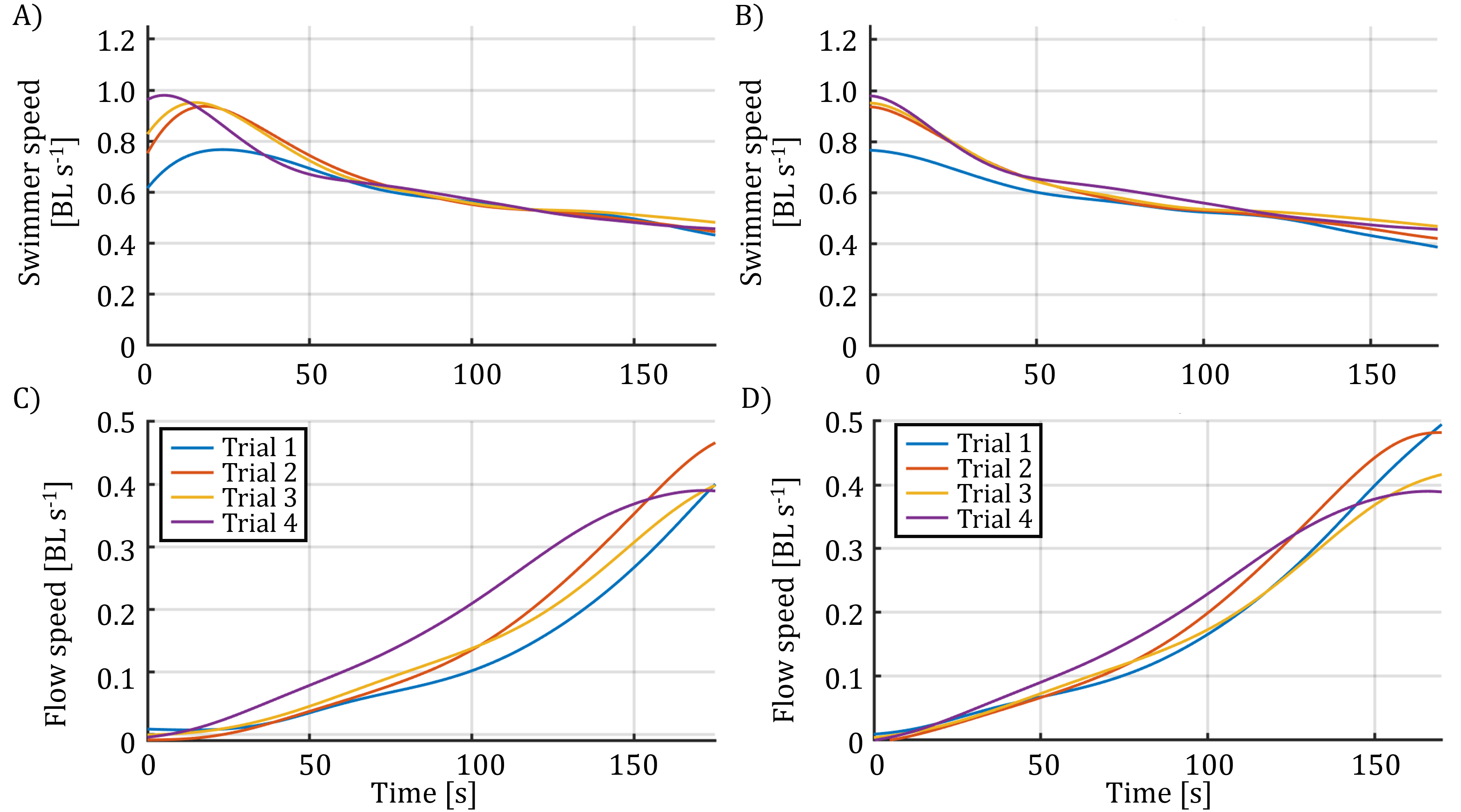}}
 \caption{Comparing unshifted and time-shifted data for the four trials completed at 17~ppt salinity. A) Original swimmer speed over time; B) time-shifted swimmer speed aligned such that peak speed occurs simultaneously; C) original peak flow speed within the aggregation over time; D) time-shift defined by (B) applied to peak flow speed.}
\label{fig:results_unshifted}
\end{figure}

Using the PIV measurement and swimmer tracks that are collected simultaneously, the swimmer-frame swimming speed was calculated for each swimmer as the swimmer's velocity in the laboratory frame minus the local fluid velocity interpolated at each swimmer's location. To visualize the consistency and variability in repeated trials, the swimmer-frame swimming speed is plotted, and although clear trends were observed in each variable, there are apparent timing differences between the trials. These timing differences may obscure direct comparisons. To address this, the data were temporally aligned by shifting each trial so that the time of peak swimmer speeds coincided (figure~\ref{fig:results_unshifted}). This alignment facilitated clearer comparisons of common trends and enabled accurate averaging across trials. Subsequent analyses and statistical comparisons were performed using this aligned data. 

Inspection of swimmer counts across salinity conditions revealed substantial variation that was not consistent across treatments (figure~\ref{fig:statistics_time}). The lowest- and highest-salinity conditions sustained roughly half as many swimmers as the intermediate conditions. Because aggregation size directly affects induced flow, this would confound any direct comparison of induced flow across salinity conditions. To statistically isolate the buoyancy contribution from these confounds, we used a multiple regression model with centerline flow velocity at the trial endpoint as the response variable, and swimmer count, swimming speed, fluid density (via salinity treatment), and the per-trial Gaussian standard deviation $\sigma_x$ of the swarm spatial distribution as predictors. The Gaussian width $\sigma_x$ was estimated from the per-timestep distribution of $x$-coordinates of detected swimmers using a normal fit (median $R^2 > 0.8$ across conditions). The multiple regression model was fit using \texttt{fitlm} in MATLAB (R2024b, MathWorks Inc., Natick, MA), which performs ordinary least squares estimation. Reported $p$-values for individual regression coefficients are from two-sided $t$-tests of the null hypothesis $\beta = 0$ using residual degrees of freedom.

\section{Theoretical framework} \label{sec:theoretical}
To interpret our experimental measurements in the context of momentum input from collective swimming, we developed an analytical framework based on the actuator-disk model~\citep{rankine_mechanical_1865}, previously applied to model collective organism swimming\citep{houghton_alleviation_2019,fu_magnetic_2023}. The model treats the aggregation as a uniform disk-shaped source of momentum, allowing the induced centerline velocity to be expressed in terms of swimmer number, swimming speed, and net buoyant force.

\subsection{Actuator-disk model and dimensional analysis}

To better contextualize experimental results and assess the physical basis of the observed variations in induced flow, we compared the experimental results directly with predictions from an actuator disk model that estimates induced flow velocities based on the conservation of momentum~\citep{rankine_mechanical_1865}. The actuator disk model assumes uniform thrust distribution across a hypothetical disk, an idealization previously applied to model collective organism swimming~\citep{houghton_alleviation_2019, fu_magnetic_2023} according to the following equation:

\begin{equation} \label{eqn:delta_w_0}
    \Delta W = \frac{-u_z}{2} + \sqrt{\frac{u_z^2 }{4} + \frac{F_b}{2 \rho A_D}},
\end{equation}

where $\Delta W$ is the induced velocity in the centerline directly behind the actuator disk, $u_z$ is the speed of the actuator disk, $A_D$ is the area of the actuator disk, $\rho$ is the density of fluid and $F_b$ is the buoyant force acting on the actuator disk. In order to translate this into collective vertical migration, a representative vertical swimming velocity will be used for $u_z$, and the buoyancy will be set equal to the net buoyant force on all swimmers, defined as:

\begin{equation} \label{eqn:F_b}
    F_b= N \Delta\rho g V_{s},
\end{equation}

where $N$ is the number of swimmers, $\Delta\rho$ is the difference in density between a swimmer and the surrounding fluid, $g$ is gravitational acceleration and $V_{s}$ is the volume of each swimmer.

Upon plugging \eqref{eqn:F_b} into \eqref{eqn:delta_w_0}, the induced flow is modeled as: 

\begin{equation} \label{eqn:delta_w_0_full}
     \Delta W = \frac{-u_z}{2} + \sqrt{\frac{u_z^2 }{4} + \frac{ N \Delta\rho g V_{s}}{2 \rho A_D}}.
\end{equation}

\subsection{Extension using a Froude number}

To explore how induced flow scales with the dominant variables (number of swimmers, density difference, and swimming speed), we defined a dimensionless\index{dimensionless} parameter, $\Lambda$, analogous to a Froude number, which compares inertial forces due to swimmer ascent speed with the buoyancy-driven momentum input from swimmer thrust. This dimensionless parameter characterizes the flow regime and determines the dominant physics:

\begin{equation}
\Lambda \equiv \frac{\rho A_D u_z^2}{2 N \Delta\rho g V_{s}}
\begin{cases}
\ll 1 & \text{[buoyancy-dominated]} \\
\gg 1 & \text{[inertia-dominated]}
\end{cases}
\end{equation}

Oceanic observations suggest that zooplankton operate primarily in transitional regimes, with $\Lambda$ values on the order of 1 to 100~\citep{kunze_biologically_2019,visser_biomixing_2007}. In the experiments presented here, $\Lambda$ is approximately unity, representing a transitional balance between inertial and buoyant forces. When $\Lambda \ll 1$, the buoyancy-driven thrust dominates over inertial forces, corresponding to a low-speed regime, analogous to a low Froude number regime. Under these conditions, the scaling relationship for induced flow simplifies to 

\begin{equation}
\Delta W \propto \sqrt{N \Delta \rho}.
\end{equation}

At high swimmer ascent speeds, $\Lambda \gg 1$, inertial forces associated with swimmer ascent speed dominate over buoyancy-driven thrust, indicating an inertia-dominated regime analogous to a high Froude number. In this limit, the swimmer-induced thrust is small relative to the background velocity squared term, and we rewrite \eqref{eqn:delta_w_0_full} as:

\begin{equation} 
    \Delta W = \frac{-u_z}{2} + \sqrt{\frac{u_z^2}{4} \left( 1 + \frac{2 N \Delta\rho g V_{s}}{\rho A_D u_z^2} \right)},
\end{equation}

and factor out the $u_z$ term:

\begin{equation} \label{eqn:factored}
    \Delta W = -\frac{u_z}{2}+ \frac{u_z}{2}\sqrt{1 + \frac{2 N \Delta\rho g V_{s}}{\rho A_D u_z^2} }.
\end{equation}

In the high-speed limit ($\Lambda \gg 1$), the swimmer thrust term is very small compared to the swimmer speed term. Therefore, we can use the binomial approximation for small $x$ in the form $\sqrt{1+x}$:

\begin{equation} 
\sqrt{1+x} \approx 1 + \frac{x}{2} \text{, when } |x|\ll1.
\end{equation}

Applying this to \eqref{eqn:factored} gives the following:

\begin{equation}
    \Delta W = -\frac{u_z}{2}+ \frac{u_z}{2} \left(1+ \frac{N \Delta\rho g V_{s}}{\rho A_D u_z^2} \right).
\end{equation}

\begin{equation}
    \Delta W =  \frac{N \Delta\rho g V_{s}}{2 \rho A_D u_z}.
\end{equation}

\begin{equation}
\Delta W \propto \frac{N \Delta \rho}{u_z}.
\end{equation}

\section{Results}

\subsection{Time evolution across salinities}
\begin{figure}
 \centering{\includegraphics[scale=0.29]{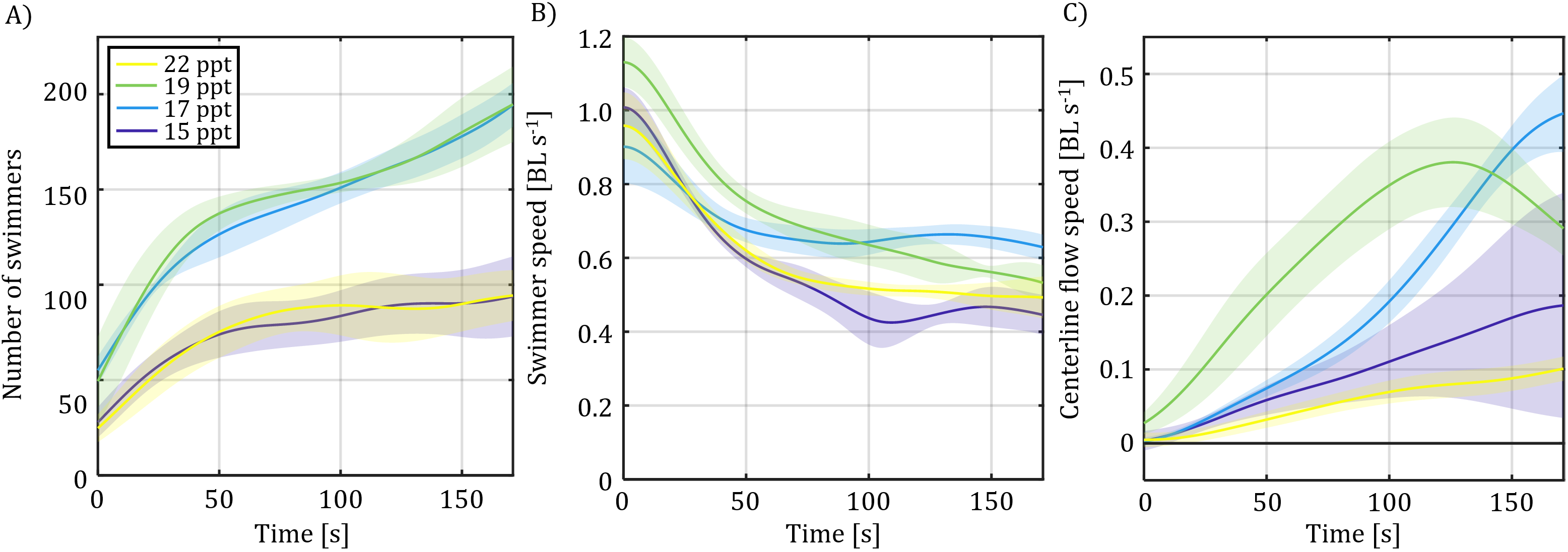}}
 \caption{Time evolution of three quantities during induced vertical migration, each averaged across four trials per salinity condition. (A) Number of swimmers in the imaging volume. (B) Swimmer-frame ascent speed. (C) Centerline flow speed of downward induced jet. Solid lines indicate mean across trials, with shaded regions representing one standard deviation. Color denotes salinity treatment (15, 17, 19, and 22~ppt). All trials were temporally aligned by shifting each so the time of peak swimmer speed coincided (see figure~\ref{fig:results_unshifted} and Methods).}
\label{fig:statistics_time}
\end{figure}

The temporal evolution of swimmer count, induced centerline flow speed, and swimmer ascent speed during each migration are shown in figure~\ref{fig:statistics_time}. Across all conditions, swimmer counts increased over time as animals migrated into the imaging volume, induced flow speeds grew correspondingly, and swimming speeds peaked early in each trial before settling to a lower steady value. The magnitude of all three quantities differed substantially across salinity treatments, therefore, pairwise comparisons across salinities cannot in isolation reveal a buoyancy effect. The linear mixed-effects regression analysis in the next subsection explicitly accounts for these covariates.

\subsection{Buoyancy effect on induced flow and regression analysis}

\begin{figure}
 \centering{\includegraphics[scale=0.24]{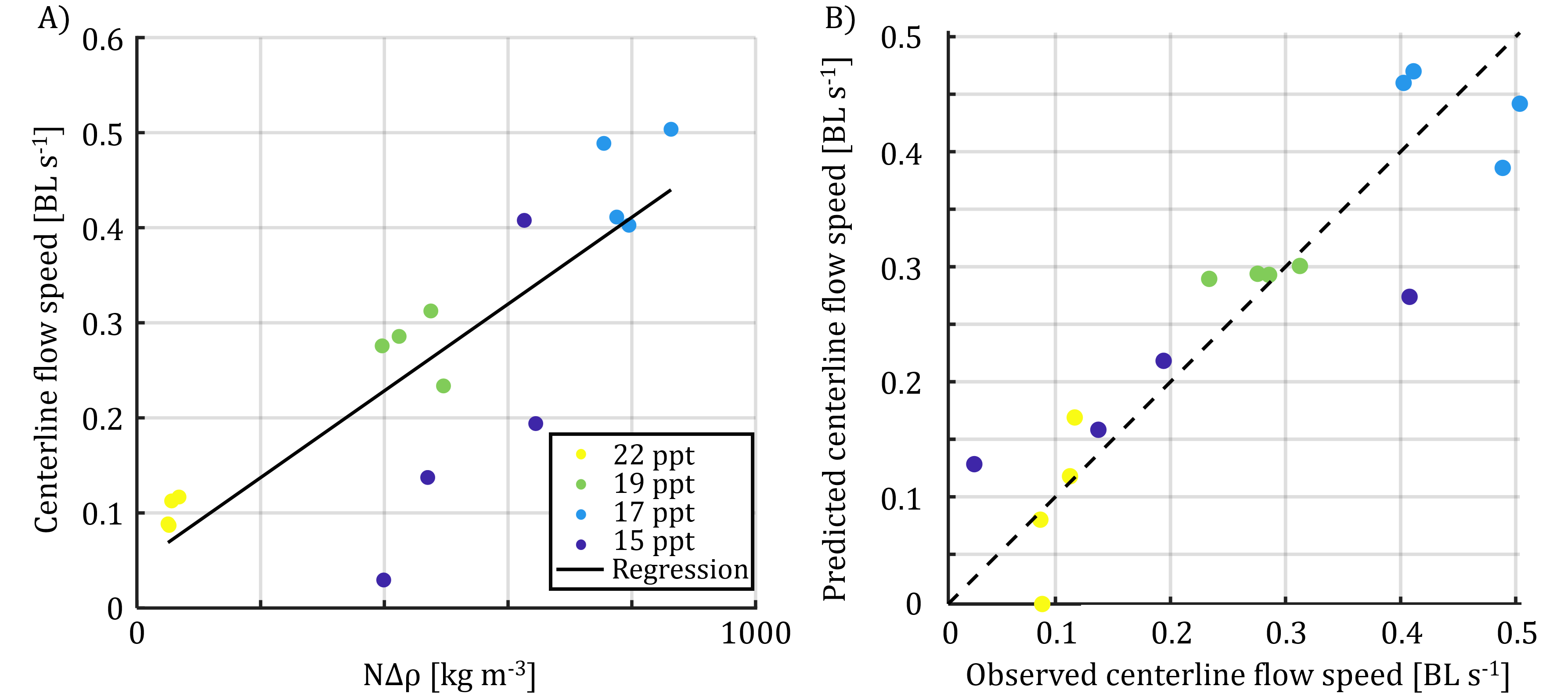}}
 \caption{Buoyancy-forcing scaling and statistical analysis of induced velocity. A) Scaling of induced velocity with buoyancy forcing, $N \Delta \rho$. Point color denotes salinity treatment (15, 17, 19, and 22~ppt). The solid line shows the best-fit linear regression ($R^2=0.70$, $p=5.9 \times 10^{-5}$), indicating that induced centerline velocity increases with buoyancy forcing. B) Observed versus predicted centerline velocities from the endpoint multiple regression including swimmer number, swimming speed, fluid density, and swarm Gaussian width $\sigma_x$ as predictors. Each point represents a single trial; point color denotes salinity treatment. The dashed line indicates the 1:1 relationship ($R^2 = 0.82$).}
\label{fig:LMEM}
\end{figure}

To evaluate whether induced velocity scales with buoyancy forcing, the measured centerline velocities from each trial were plotted against the forcing term $N \Delta \rho$ (figure~\ref{fig:LMEM}A), where $N$ is the number of swimmers and $\Delta \rho = \rho_s-\rho$ is the density difference between the swimmer and the surrounding fluid. Linear regression revealed a strong positive relationship between induced velocity and buoyancy forcing ($R^2=0.70$, $p=5.9 \times 10^{-5}$).

To further account for variation in swarm size, swimming kinematics, and swarm spatial organization, a multiple regression analysis including swimmer count, swimmer-frame swimming speed, fluid density, and swarm Gaussian width $\sigma_x$ was performed on the trial endpoint values. Fluid density was the only statistically significant individual predictor of induced flow at $\alpha = 0.05$ ($\beta = -3.05 \times 10^{-2}$, SE $= 1.01 \times 10^{-2}$, $p = 0.012$). Swimmer-frame swimming speed had the largest raw partial coefficient ($\beta = 0.73$, SE $= 0.36$) but reached only marginal significance ($p = 0.065$), suggesting a real behavioral contribution. The overall model explained 82\% of the variance in flow speeds ($R^2 = 0.82$, $F$-test $p = 4 \times 10^{-4}$, $n = 16$ trials); the agreement between observed and predicted endpoint velocities (figure~\ref{fig:LMEM}B) shows that the four predictors together capture most of the observed variation in induced flow.

\subsection{Comparison with actuator-disk model predictions}

\begin{figure}
 \centering{\includegraphics[scale=0.29]{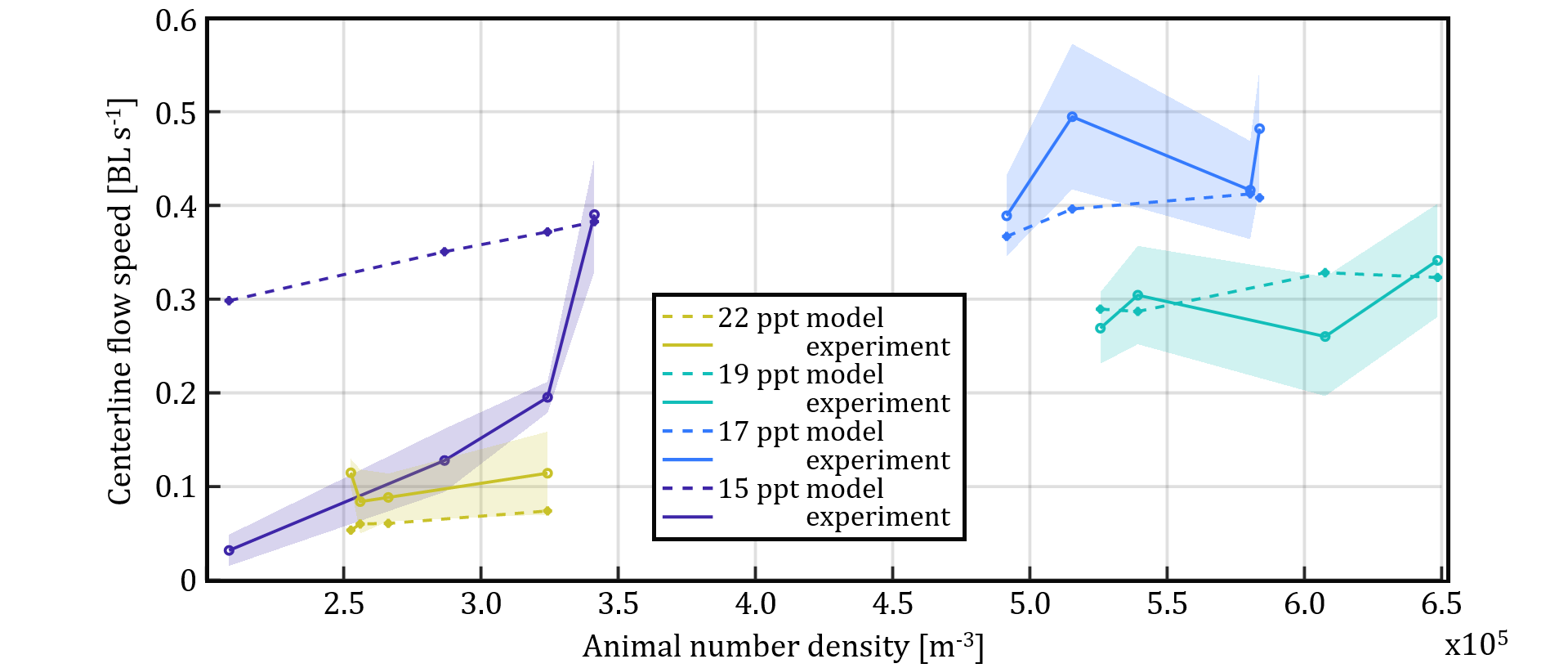}}
 \caption{Measured induced centerline flow velocities at final recorded time versus animal number density for each salinity condition (15, 17, 19, and 22~ppt). Each data point represents an individual trial, with shaded regions indicating the standard deviation of measured flow speeds at 170 s. The dotted line denotes the prediction from the actuator disk model.}
\label{fig:act_disk}
\end{figure}

Using the theoretical framework developed in the previous section, estimated induced centerline velocities are compared to experimental measurements across all four salinity conditions (figure~\ref{fig:act_disk}). Predictions were computed for each trial individually using the trial-specific number of swimmers, ascent speed, and fluid density.

The three-dimensional organization of swimmers and their induced flow structures is quantified in the supplementary information. Based on these results, we assume that the spread of swimmers is axisymmetric about the center of the tank and Gaussian in spread. The number of swimmers in a 2-D sample is used to estimate the 3-D animal number density. Additional variables defined are listed in table \ref{tab:disk_values}. 
The actuator disk model captures the first-order dependence of induced flow on swimmer number and buoyancy forcing, but over-predicts the 15~ppt condition. This discrepancy indicates additional behavioral or hydrodynamic effects not captured by the simplified model. Since the ideal salinity for brine shrimp is 35~ppt, it is possible that swimming in 15~ppt water altered swimmer condition or swimming kinematics. This discrepancy, along with non-linearities inherent to the swimmer-flow coupling, is examined further in the Discussion.

\begin{table}
  \centering
  \begin{tabular}{cccc}
     \toprule
      Symbol & Variable & Value & Unit\\
    \midrule
      $u_z$  & swimming velocity & taken from data & cm s$^{-1}$ \\
      $g$  & gravitational acceleration  & 9.8 & m s$^{-2}$ \\
      $V_s$  & swimmer volume & $0.1$  & cm$^3$ \\
      $\rho_s$  & swimmer density & 1015 & kg m$^{-3}$ \\
      $\rho$  & seawater density  & taken from data & kg m$^{-3}$ \\
      $A_D$& actuator disk area & 20 & cm$^2$ \\
    \bottomrule
  \end{tabular}
  \caption{Variables used in theoretical actuator disk model for induced flow in vertical migration.}
  \label{tab:disk_values}
\end{table}

All trials for the 15~ppt condition fall in the buoyancy-dominated regime, as defined by $\Lambda$. Therefore, although the median swimmer speed used in model predictions did not fully capture this skewed velocity distribution, the magnitude of the change would be small relative to the impact of the number of swimmers and buoyant forces, and if it explicitly accounted for slower swimmers, would theoretically yield even higher predictions of induced velocity. This outcome arises from the inverse relationship between the induced flow velocity and the swimming ascent speed described by the actuator disk model scaling. However, this unexpected mismatch underscores a critical limitation in the actuator disk analogy as applied here: the model assumes uniform momentum transfer across the disk area, neglecting variations in wake interactions that occur due to swimmer velocity differences. Swimmers moving at higher ascent speeds produce stronger, more coherent wakes, potentially increasing local wake interactions and altering the scaling relationship between swimmer speed and induced flow velocity. Thus, the assumption of uniform momentum distribution fails to capture the increased complexity of hydrodynamic interactions within the collective migration, particularly in cases where swimmer velocity distributions are skewed or nonuniform. 

\section{Discussion}

We quantified the induced flow generated during collective vertical migration using simultaneous swimmer tracking and particle image velocimetry while varying fluid density to alter swimmer buoyancy. Induced centerline velocity was found to be significantly associated with fluid density even after accounting for variation in swimmer number, speed, and distribution. Thus, within the limits of this experimental system, the results support a relationship between buoyancy-related forcing and aggregation-scale induced flow. At the same time, because buoyancy was manipulated indirectly through salinity, behavioral responses to salinity cannot be excluded. The present results should therefore be interpreted as evidence of buoyancy-modulated flow generation rather than as a strict isolation of buoyancy from all biological responses.

The scaling analysis provides further support for this interpretation. When the induced velocity was plotted against the buoyancy-forcing term $N\Delta\rho$, the data showed a strong positive trend, consistent with the actuator-disk prediction that a greater net buoyant load requires greater thrust and therefore generates a stronger counterflow. The 15~ppt condition deviated from the main trend and from the actuator-disk predictions. Rather than treating this case as noise, we interpret it as evidence of an important limitation of the simplified model. The actuator-disk framework captures the first-order dependence of induced flow on swarm-scale forcing, but it does not include potential changes in swimming behavior, body orientation, wake interaction, or velocity distributions that may emerge under strongly altered salinity conditions.

This limitation is especially apparent in the 15~ppt trials, where swimmer speed distributions were more skewed and many individuals swam substantially more slowly than the representative median values used in the model. This suggests that collective hydrodynamic output may depend not only on the mean swimming speed of the aggregation but also on the distribution of individual swimmer velocities within it. This point is important beyond the present study, because many models of collective biological flows rely on representative mean quantities that may obscure biologically meaningful variability. Future work should therefore test how sensitive these predictions are to swimmer heterogeneity, and salinity-dependent behavior, and should evaluate the scaling directly in field or mesocosm settings where density gradients and collective migration occur simultaneously.

The theoretical model remains useful because it provides a compact framework for organizing the experimental observations and extending them to oceanographically relevant regimes. The dimensionless parameter

\begin{equation}
\Lambda = \frac{\rho A_D u_z^2}{2 N \Delta\rho g V_{s}},
\end{equation}

compares inertial swimming effects with buoyancy-driven thrust and therefore determines the form of the induced-flow scaling. In the buoyancy-dominated limit ($\Lambda\ll1$), the model predicts a square-root dependence on buoyancy forcing, $\sqrt{N\Delta\rho}$, whereas in the inertia-dominated limit the dependence becomes approximately linear in $N\Delta\rho$.

\begin{figure}
 \centering{\includegraphics[scale=0.225]{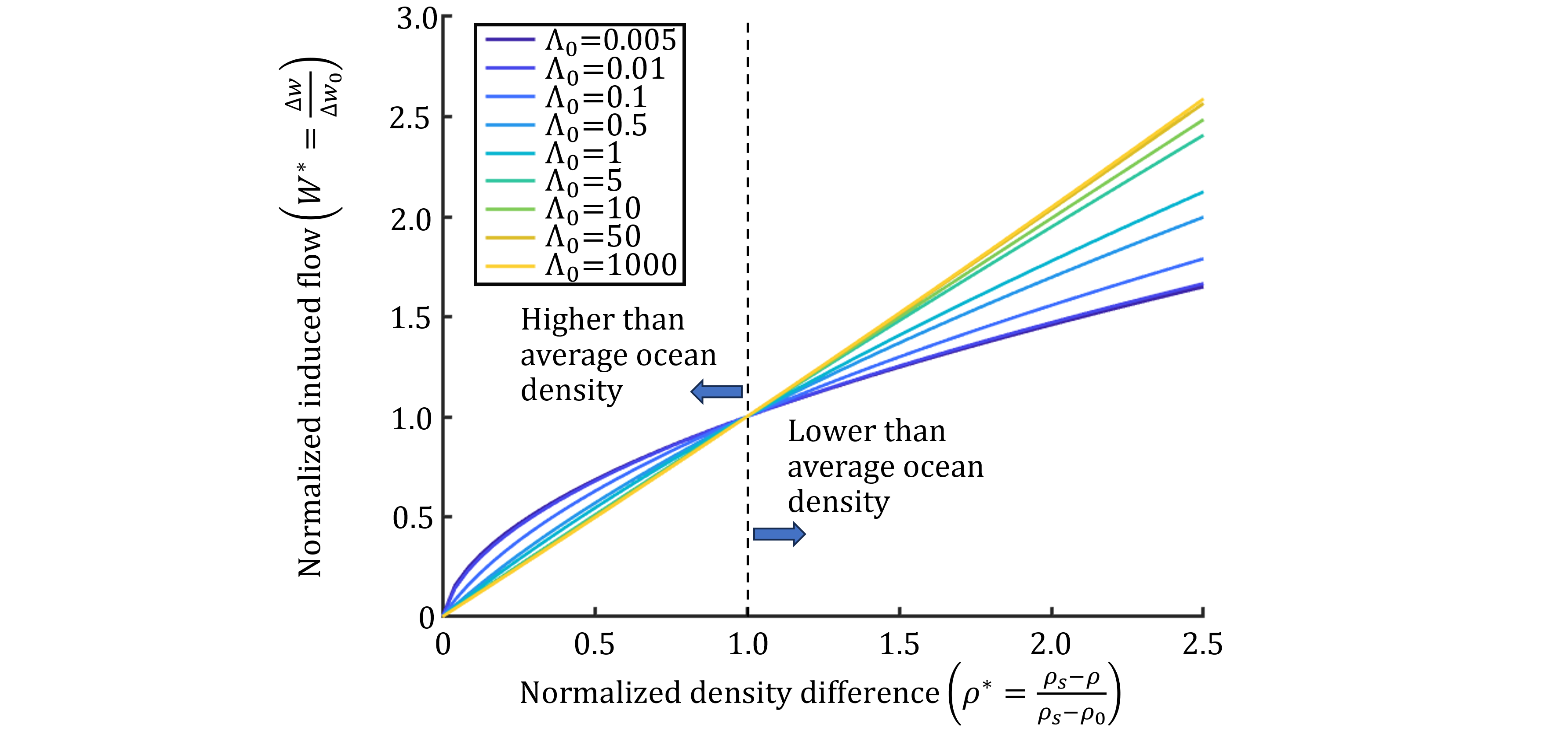}}
 \caption{Normalized induced flow velocity, $W^*$, is plotted against the normalized density difference, $\rho^*$, illustrating how induced flow velocity changes across typical oceanographic density conditions. Curves are labeled with different values of the dimensionless parameter $\Lambda$, representing the relative importance of inertial versus buoyant forces. The left region ($\rho^*<1$) indicates fluid denser than the average ocean reference, resulting in reduced induced flows ($W^*<1$). The right region ($\rho^*>1$) corresponds to lighter fluid, enhancing induced flows ($W^*>1$). Curves transition from buoyancy-dominated regimes ($\Lambda \ll 1$), scaling approximately as $(\rho^*)^{1/2}$, to momentum-dominated regimes ($\Lambda \gg 1$), exhibiting linear scaling with density difference.}
\label{fig:non_dim_lambda}
\end{figure}

To extend the results beyond the laboratory, we defined a normalized density-difference parameter, $\rho^*$, relative to a reference ocean density, $\rho_0 = 1036$~kg~m$^{-3}$, with swimmer density set to $\rho_s = 1060$~kg~m$^{-3}$ to represent a negatively buoyant organism. The corresponding normalized induced velocity, $W^*$, shows how environmentally realistic changes in ambient density alter the induced flow predicted by the model (figure~\ref{fig:non_dim_lambda}). At $\rho=\rho_0$, the normalized induced flow is unity by definition. As ambient density increases toward swimmer density, the swimmer approaches neutral buoyancy and the induced flow decreases toward zero. Conversely, as ambient density decreases below the reference value, induced flow increases. Across realistic ocean density ranges, the model predicts that this enhancement is substantial but bounded, reaching at most roughly 2.5 times the reference value.

These predictions have important implications for interpreting field measurements of collective vertical migration in the ocean. Measurements obtained in denser water masses may underestimate the induced flows that would occur under average ocean conditions, whereas measurements from lighter-density environments may overestimate mixing potential if extrapolated without accounting for buoyancy effects. More broadly, the results suggest that environmental structure can influence the hydrodynamic consequences of collective motion, not only through its effects on where animals aggregate, but also through how strongly their collective swimming couples momentum into the surrounding fluid. This is especially relevant for vertically migrating organisms that routinely traverse stratified water columns and experience systematic changes in ambient density over depth.

Overall, this study shows that environmentally modulated buoyancy can alter the strength of flows generated by collective vertical migration. By combining swimmer-resolved measurements, flow measurements, statistical control for confounding variables, and simple scaling analysis, the results support a mechanistic link between buoyancy forcing and aggregation-scale induced flow. More broadly, they highlight how collective motion in complex environments can produce fluid dynamical consequences that depend on the physical properties of the surrounding medium.


\section*{Acknowledgments}
The authors gratefully acknowledge M. K. Fu (California Institute of Technology) for advice on both experimental protocols and theoretical model implementation. The authors would like to thank the reviewers from the Integrative and Comparative Biology for their detailed feedback and helpful suggestions that significantly improved the presentation of this work.

\section*{Funding}
This work was supported by funding from the US National Science Foundation through the Alan T. Waterman Award and Graduate Research Fellowship Grant No. DGE 1745301.

\section*{Competing interests}
The authors declare no competing interests.

\section*{Data availability}
All data generated and discussed in this study are available within the article and its supplementary files, or are available from the authors upon request.

\section*{Supplementary material}
Supplementary material will be available upon publication.

\bibliographystyle{plainnat}   
\bibliography{references}   

\newpage

\section{Supplementary information}

\subsection{Experimental configurations}
Two experimental configurations were used (Table~\ref{tab:exp_compare}). High-resolution measurements (A) used approximately 6000 animals to ensure sufficient sampling density for the 3-D characterization of swimmer trajectories and induced flow structure. Long-duration measurements (B) used approximately 3000 animals to avoid optical saturation and tracking overlap during the extended recordings used for the time-resolved analyses in the main text. Results from setup (A) are contained to the supplement.

\begin{table}[ht]
  \centering
  \begin{tabular}{cccccc}
    \toprule
     &
    \makecell{Recording\\ length (s)} &
    \makecell{Capture\\ rate (fps)} &
    \makecell{Depth, $y$\\ (BL)} &
    \makecell{Scanning\\ rate (Hz)} &
    \makecell{Swimmers\\ added} \\
    \midrule
    A & 40  & 2000 & 11   & 5 & 6000 \\
    B & 170 & 60   & 2.5  & 5 & 3000 \\
    \bottomrule
  \end{tabular}
  \caption{Settings used for the two 3-D scanning experiments. Method A has higher spatial and temporal resolution; method B has a recording length five times longer.}
  \label{tab:exp_compare}
\end{table}

\subsection{Salinity and density profile verification}

\begin{figure}
 \centering{\includegraphics[scale=0.225]{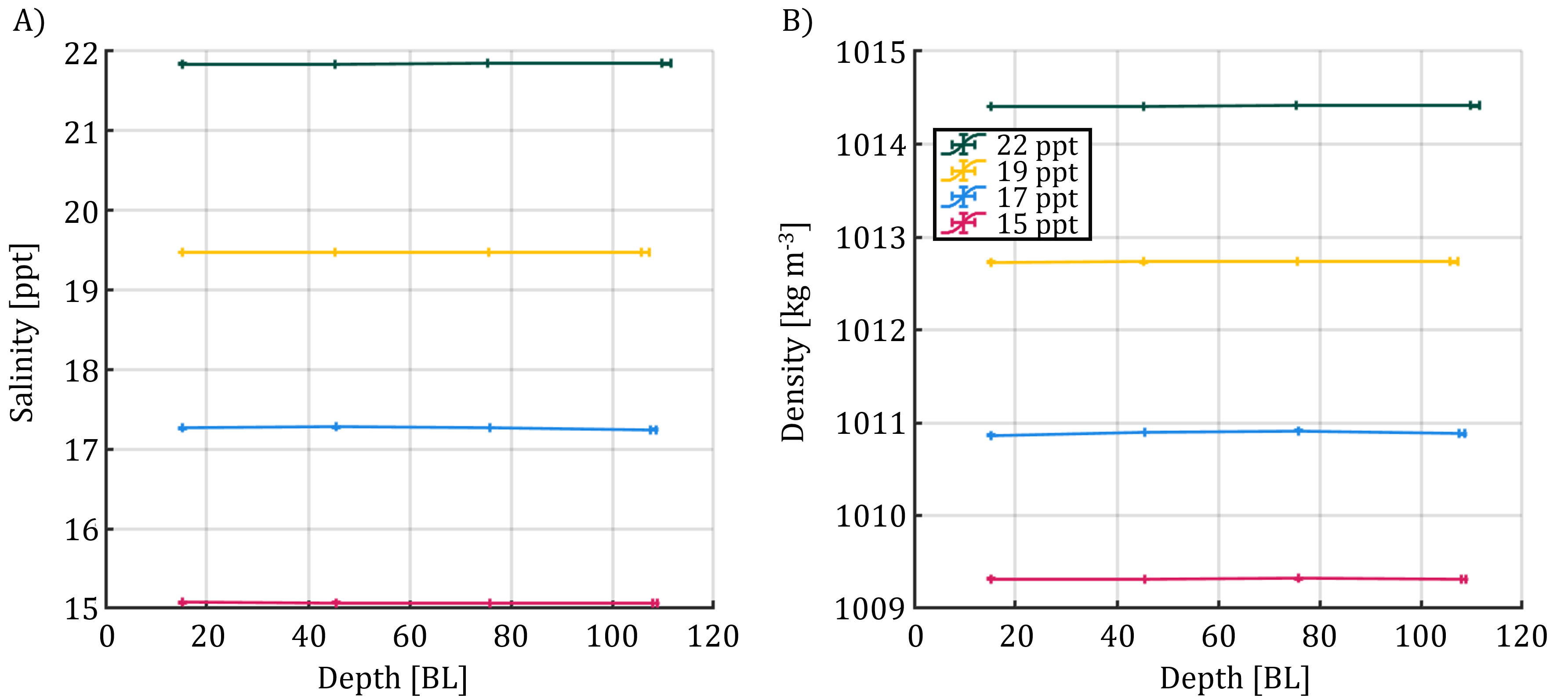}}
 \caption{Measured A) salinity and B) density values for the four experimental conditions vs.\ tank depth (in body lengths, $1~\text{BL} = 1$~cm). Salinity was adjusted to four target values (15, 17, 19, and 22~ppt) and verified using a CastAway CTD profiler (SonTek, USA), with five measurements taken from different locations in the tank immediately before loading the animals. Values are shown as the average $\pm$ (accuracy $+$ standard deviation) across the five locations.}
\label{fig:CTD}
\end{figure}

Tank salinity and density were measured immediately before each migration trial using a CastAway CTD profiler (SonTek, USA), with five vertical profiles taken at the four corners and center of the tank (figure~\ref{fig:CTD}). The profiles confirmed that the prescribed salinity was uniform with depth in all four conditions, and the inter-location standard deviations of the depth-averaged values were comparable to the resolution of the instrument.

\subsection{Robustness of buoyancy effect to swarm spatial organization}

The endpoint multiple regression analysis in §3.2 includes the per-trial Gaussian standard deviation $\sigma_x$ of swimmer $x$-positions as a covariate to test whether differences in swarm spatial organization could account for the observed flow differences across salinities. Across all four salinity conditions, the per-timestep Gaussian fit was strongly supported (median $R^2 > 0.8$), justifying treatment of $\sigma_x$ as a meaningful organization metric. At the trial endpoint with $n = 16$ trials, $\sigma_x$ was not a significant individual predictor ($p = 0.40$, see §3.2), and density's effect on flow remained significant after this control.

\begin{figure}
 \centering{\includegraphics[scale=0.5]{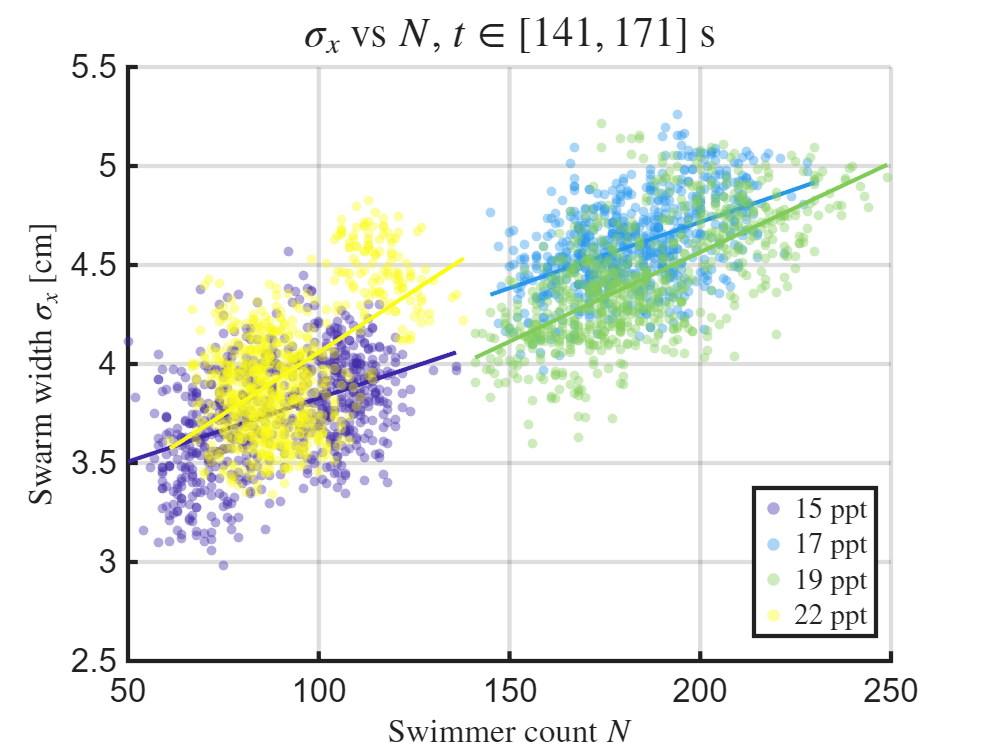}}
 \caption{Per-timestep Gaussian width $\sigma_x$ of the swarm as a function of swimmer count $N$, pooled over a steady-state time window. Color denotes salinity treatment (15, 17, 19, 22~ppt); lines are per-salinity linear fits. The width increases with $N$ at all salinities, with somewhat more compact swarms at 15~ppt at high $N$.}
\label{fig:supp_sigma}
\end{figure}

\subsection{Results from Protocol A}

\subsubsection{Three-dimensional swarm and flow structure}\label{sec:3D_results}

The three-dimensional swimmer trajectories collected with the high-resolution configuration A were used to characterize the spatial organization of the swarm and the structure of the induced flow. Swimmer positions in the horizontal ($x$-$y$) plane, sampled at the temporal midpoint of each tracked trajectory, were strongly aligned with a Gaussian distribution; Kolmogorov-Smirnov tests against the alternatives lognormal, gamma, Weibull, and Nakagami rejected each at $p<0.05$. A 2-D Gaussian fit to the swimmer-position histogram yielded spatial standard deviations $\sigma_x \approx 2.93$~BL and $\sigma_y \approx 2.54$~BL with a modest center offset ($x_0 \approx 0.5$~BL, $y_0 \approx 0$~BL).

The time-averaged vertical-velocity field was also well described by a 2-D Gaussian (goodness-of-fit $R^2 > 0.95$), with widths $\sigma_x \approx 3.16$~BL and $\sigma_y \approx 2.78$~BL. The similarity between the swimmer-position and induced-flow distributions supports treating the swarm as a localized, near-axisymmetric source of momentum. We define a characteristic jet width $\sigma_{jet} = 2.97$~BL and centerline speed $w_{jet} = 0.23$~BL~s$^{-1}$, yielding a jet Reynolds number

\begin{equation}
\text{Re}_{jet} = \frac{w_{jet}\, \sigma_{jet}}{\nu} \approx 66
\end{equation}

at $20^{\circ}$C, where $\nu = 1.04 \times 10^{-2}$~cm$^{2}$~s$^{-1}$ is the kinematic viscosity of 20~ppt salt water. This value places the induced jet within the intermediate-Reynolds regime ($1 < \text{Re} < 1000$) characteristic of many planktonic swimmers.

\subsubsection{Spatial distribution of swimmers}

\begin{figure}
 \centering{\includegraphics[scale=0.275]{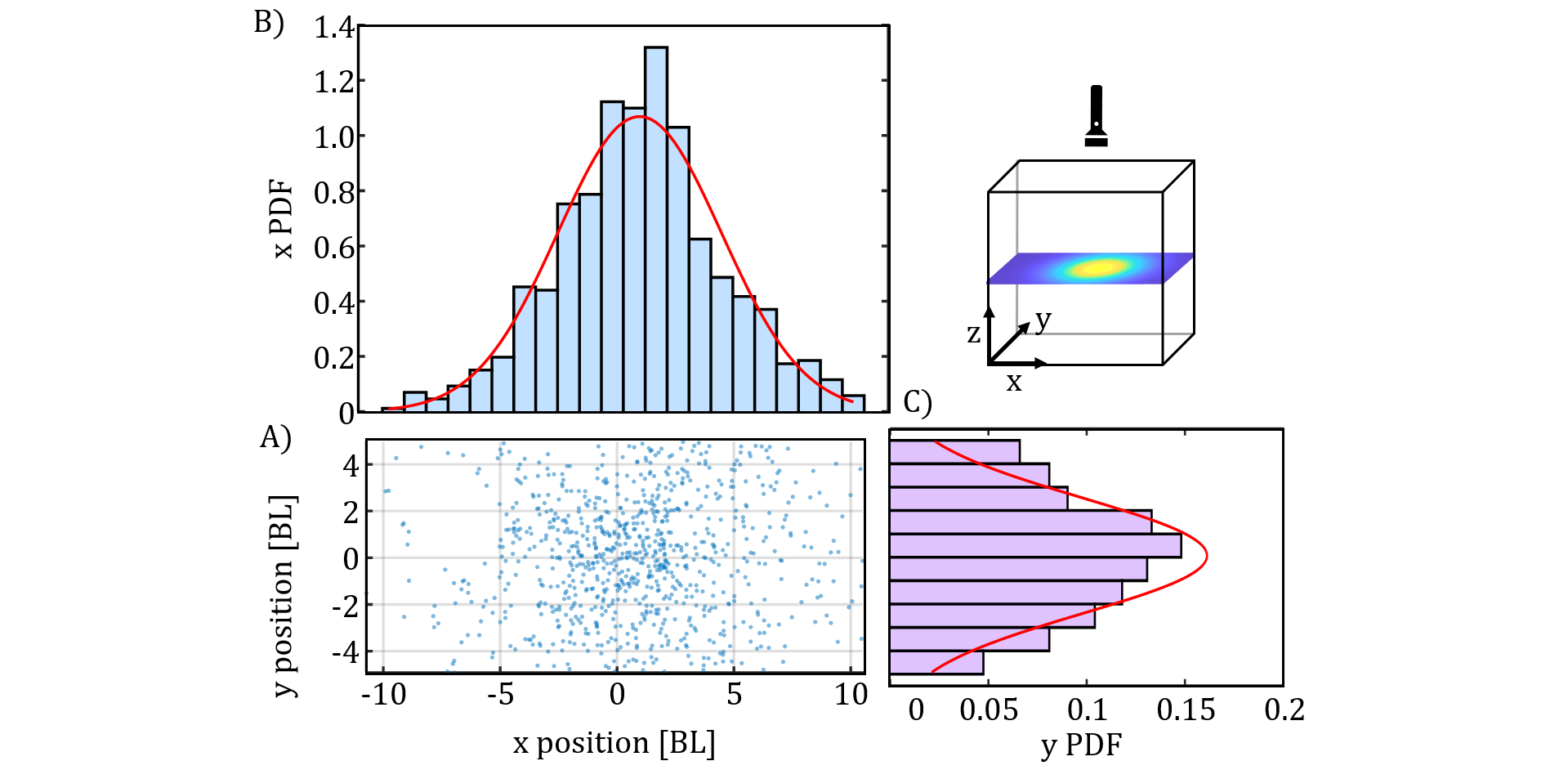}}
 \caption{A) Scatter plot illustrating swimmer positions in the horizontal ($x$-$y$) plane, representing midpoint locations of measured trajectories. Histograms depict the probability density functions (PDFs) of B) $x$ and C) $y$ swimmer positions along with the corresponding Gaussian fits used for K–S statistical testing. The inset orientation diagram shows the tank volume (22~cm × 22~cm × 10~cm in $x$, $z$, and $y$, respectively), with the target flashlight at positive $z$ approximately centered in the $x$-$y$ plane.}
\label{fig:results_positions_spread}
\end{figure}

Using the 3-D swimmer trajectories collected with setup A in table \ref{tab:exp_compare}, the spatial distribution of the swimmers was examined in the horizontal ($x$-$y$) plane. Since swimmer trajectories vary in their start times, end times, and overall durations, location data were extracted from different timesteps by using the $x$ and $y$ position of each swimmer at the middle recorded time point of each recorded trajectory. This location was selected as representative because simple time averaging or including all tracked points could bias the results toward slower swimmers or areas with increased background flow. The highest concentration of swimmers was found in the center of the tank, coinciding with the position of the target flashlight (figure~\ref{fig:results_positions_spread}).

To statistically assess whether the distribution of swimmer locations conforms to a known theoretical distribution, a Kolmogorov-Smirnov (KS) test~\citep{smirnov_table_1948} was used. The KS statistic is defined as the maximum distance between the empirical cumulative distribution function (CDF) and the CDF predicted by the theoretical distribution. The horizontal ($x$) and depth ($y$) distributions were first examined independently. Distributions of the swimmers exhibited strong statistical alignment with the Gaussian distribution, as indicated by the KS statistics of 0.0392 and 0.0347, and the p values of 0.1376 and 0.2421, respectively (figure~\ref{fig:results_positions_spread}). Alternative distributions—namely, lognormal, gamma, Weibull, and Nakagami—were rejected (p < 0.05), further reinforcing the Gaussian distribution to characterize swimmer distribution within the tank cross-section.

The fitted Gaussian distributions produced spatial standard deviations in $x$ and $y$ of $\sigma_x = 3.73$~cm and $\sigma_y =2.47$~cm respectively, along with the corresponding center positions $x_0 = 0.74$~cm and $y_0 = 0.07$~cm, suggesting an approximately 50\% greater horizontal spread and moderate offset from the $x$-axis center. However, performing separate fit analyses in each dimension may overstate the perceived anisotropy by overlooking correlations between the $x$ and $y$ coordinates. Additionally, the reduced depth dimension in $y$ results in limited sampling at the tails. To more accurately assess the anisotropy, the swimmer location data were fitted to a 2-dimensional Gaussian function of the form:

\begin{equation} \label{eq:fit_gaus}
    f(x,y)= a e^{-\left(\frac{(x-x_{0})^2}{2 \sigma_x^2}+\frac{(y-y_{0})^2}{2 \sigma_y^2} \right)}+b.
\end{equation}

\begin{figure}
 \centering{\includegraphics[scale=0.275]{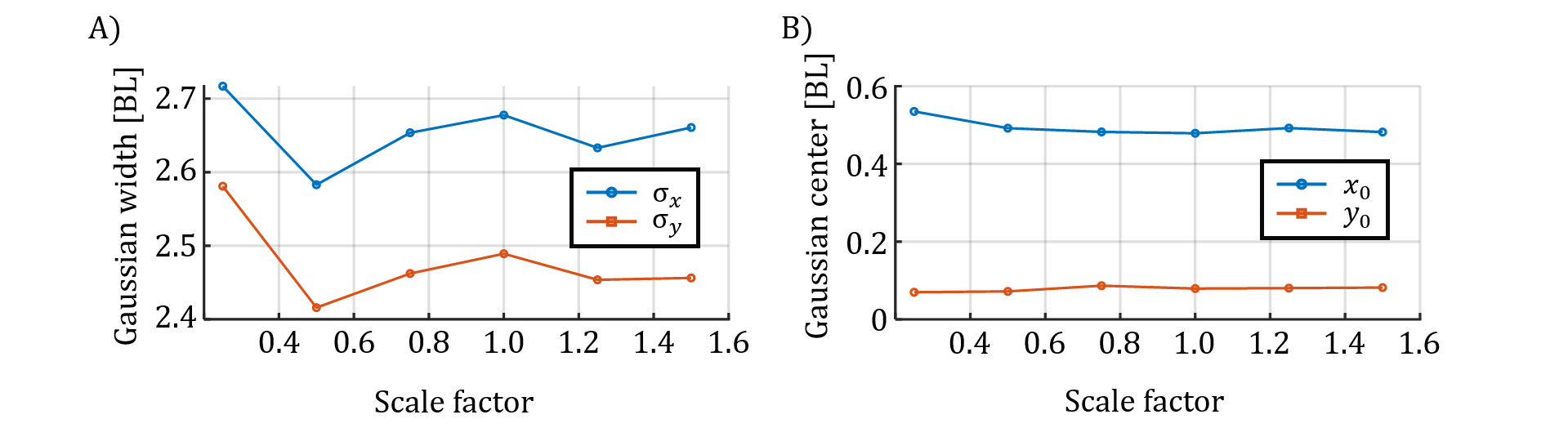}}
 \caption{Swimmer distribution 2-dimensional Gaussian fit parameters to varying bin resolutions: A) Gaussian widths across scale factors; B) Gaussian distribution centers across scale factors. Displays low sensitivity to binning.}
\label{fig:results_swimmer_bins}
\end{figure}

To ensure robustness against spatial binning, the fitting process was repeated across a range of grid resolutions, from 0.25× to 1.5× the PIV data resolution (63 × 17 bins in $x$ and $y$, respectively). The results demonstrated stability in the Gaussian width parameters (average values: $\sigma_x = 2.93$~cm,  $\sigma_y = 2.54$~cm) and the distribution center (average values: $x_0 = 0.54$~cm,  $y_0 = 0.06$~cm), with variations of $\pm 6\%$ (figure~\ref{fig:results_swimmer_bins}). However, bin resolution significantly affected swimmer counts per bin, affecting the amplitude ($a$) and baseline parameters ($b$), which were subsequently excluded from analysis.

A modest anisotropy of approximately 8\% between $\sigma_x$ and $\sigma_y$ may be due to the difference in the sampling ranges between $x$ and $y$ or may reflect the behavior of the swimmer. Although brine shrimp show heightened sensitivity to blue and shorter wavelengths, they still maintain some minimal sensitivity to red light and longer wavelengths. Consequently, the red laser sheet cutting across the tank along the x-axis could have led to a slightly wider distribution of swimmers along this axis.

Based on the 2-D Gaussian fit, we assume the swimmer distribution is axisymmetric about the tank center and approximately Gaussian in spread. The number of swimmers in $k$ 2-D sampled annular areas, approximated as,

\begin{equation}
    \Delta A_k = 2 \Delta y (r_{k+1} - r_k),
\end{equation}

where $\Delta y$ is the depth scanned, 2~cm. Each area has a different animal number density, $n_k$, which is calculated as:

\begin{equation}
    n_k= \frac{N_k}{\Delta A_k} \text{ [swimmers BL$^{-2}$]} .
\end{equation}

This animal number density is then multiplied by the full area of each annular region to get the final number of swimmers:

\begin{equation}
    N= \sum_k n_k \pi (r^2_{k+1} - r^2_k) .
\end{equation}

This estimate is used in the actuator-disk model predictions of the main text.

\subsubsection{Induced flow structure}

The vertical, $z$, component of the fluid velocity, $w$, was measured using PIV simultaneously with the swimmer trajectories. To assess the temporal and vertical ($z$) stability of the flow, the ratio of the standard deviation to the mean velocity at four representative points to calculate the coefficient of variation (CoV). The maximum CoV value over time was 10\%, indicating low temporal fluctuations, suggesting that the flow can be treated as steady. The maximum vertical ($z$) variability was 23\%, reflecting low to moderate vertical ($z$ dimension) variability. Consequently, the vertical velocity, $w$, field may be represented by its time-averaged and vertically-averaged vertical velocity distribution within the horizontal ($x$-$y$) plane.

\begin{figure}
 \centering{\includegraphics[scale=0.275]{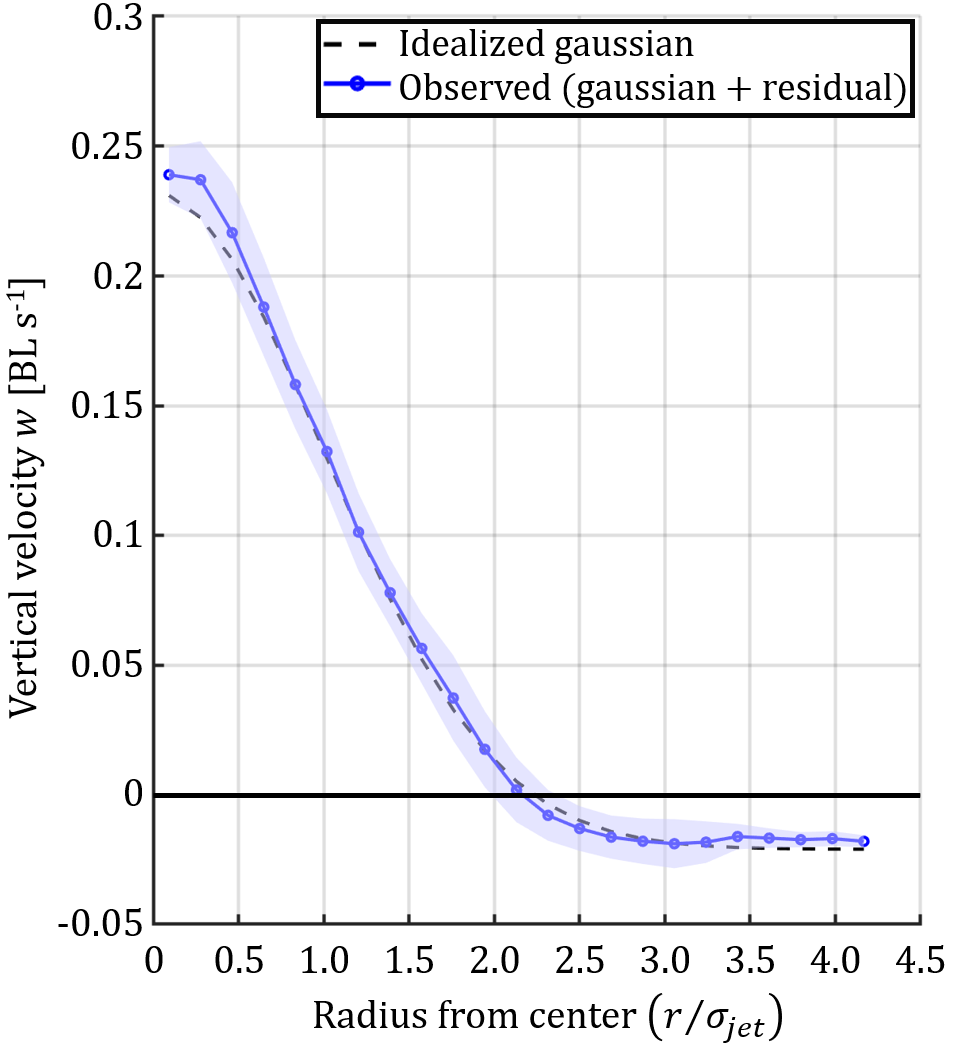}}
 \caption{Radial distribution of velocity magnitude from the tank center, comparing an idealized Gaussian profile based on 2D-fit values and the observed velocity distribution (ideal Gaussian plus mean residual with shaded standard deviation).}
\label{fig:gaus_fit}
\end{figure}

The averaged velocity distribution exhibited a distinctly Gaussian-like shape, suggesting an axisymmetric jet-like structure (figure~\ref{fig:gaus_fit}). The mean velocity distribution was fit using the two-dimensional Gaussian function \eqref{eq:fit_gaus} to quantify this observation. Again, to evaluate the stability of this Gaussian fit, spatial bin resolution was systematically varied from 0.25$\times$ to 1.5$\times$ the original PIV data resolution (63 $\times$ 17 bins for $x$ and $y$, respectively). 

\begin{figure}
 \centering{\includegraphics[scale=0.275]{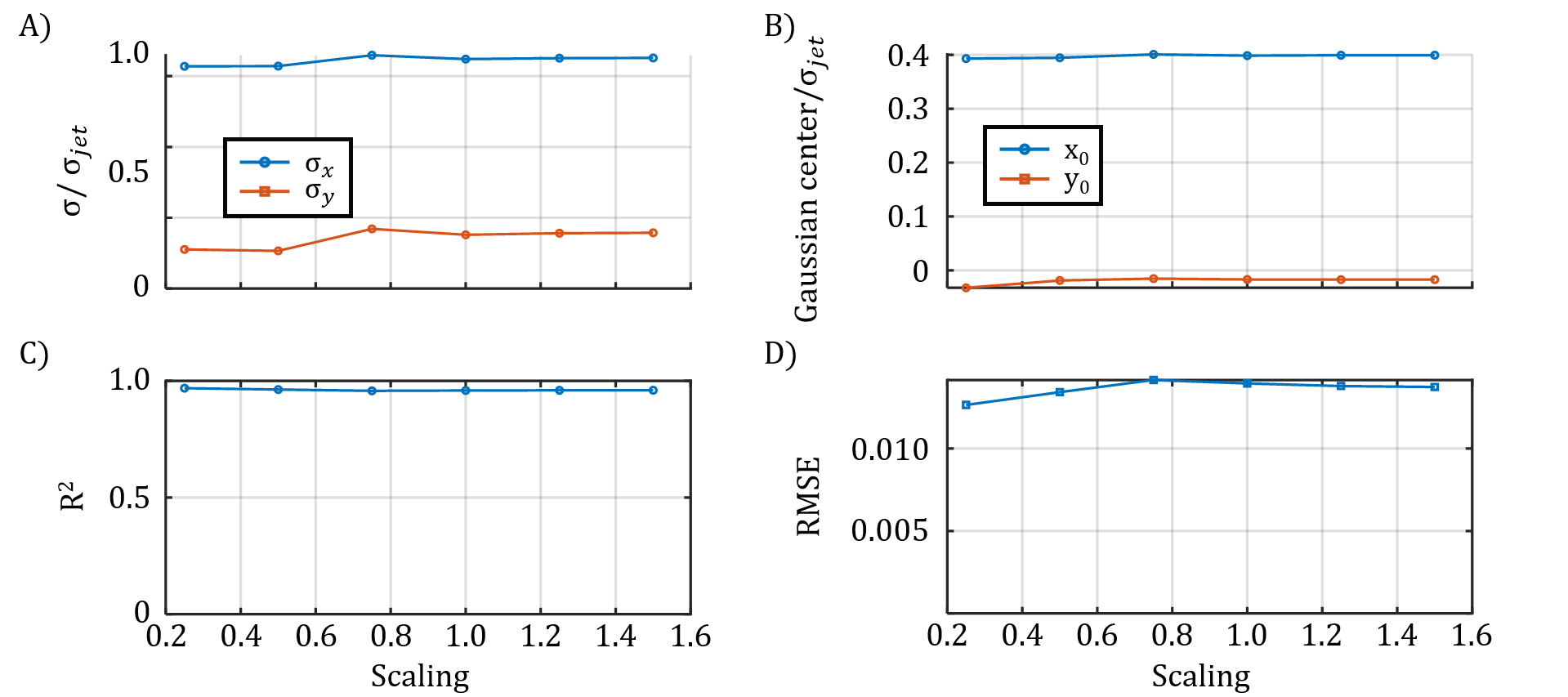}}
 \caption{Stability and statistical robustness of the Gaussian fit across bin resolutions. A) Stability of Gaussian widths: $\sigma_x$, $\sigma_y$. B) Stability of Gaussian distribution center coordinates: $x_0$, $y_0$. C) Goodness-of-fit metrics $R^2$. D) Root mean square error.}
\label{fig:flow_bins}
\end{figure}

The parameters obtained from the Gaussian fit, including the spread widths ($\sigma_x$, $\sigma_y$) and the coordinates of the distribution centers ($x_0$, $y_0$), demonstrated stability, with variations less than 2\% across all scales tested (figure~\ref{fig:flow_bins}). Furthermore, goodness-of-fit metrics were consistently strong, with $R^2$ values greater than 0.95, and the root mean square error (RMSE) ranging from 0.012 to 0.014. 

The Gaussian velocity fit (figure~\ref{fig:gaus_fit}) demonstrated a slight elliptical anisotropy ($\sigma_x=3.16$~cm, $\sigma_y=2.78$~cm) and a center offset ($x_0=1.19$~cm, $y_0=-0.05$~cm). The similarity in distribution characteristics between swimmer positions and the induced flow, particularly this slight anisotropy, indicates a direct connection through the swimmers acting as localized momentum sources. 

For subsequent analysis, we define an average Gaussian width, $\sigma_{jet} = 2.97$~cm, which is the mean of $\sigma_x$ and $\sigma_y$ from the Gaussian fit with the lowest total residual. This value serves as the characteristic normalization length scale. The centerline flow speed derived from this optimal fit ($a+b$) is designated as the characteristic jet centerline flow speed, $w_{jet} = 0.23$~cm~s$^{-1}$.

Using this length and speed scale, we calculated the jet Reynolds number:

\begin{equation}
    \text{Re}_{jet} = \frac{w_{jet}\sigma_{jet}}{\nu}= 66
\end{equation}

where $\nu$ is the kinematic viscosity of 20~parts~per~thousand salt water at 68$^{\circ}$F (20$^{\circ}$C), $1.04 \times 10^{-2}$~cm$^2$ s$^{-1}$. This calculation places the induced jet within the intermediate Reynolds number regime ($1<Re<1000$). In this regime, inertial effects become significant enough to induce some coherent vortices and unsteadiness, while viscous effects continue to maintain clearly defined spatial gradients.

To further characterize the induced flow structure, we derived the magnitude of shear\index{shear} rate from the spatial gradient of the vertical velocity field. Specifically, the instantaneous shear magnitude was computed using the following formula: 

 \begin{equation}
     |\nabla w| = \sqrt{\left(\frac{\partial w}{\partial x} \right)^2 + \left(\frac{\partial w}{\partial y} \right)^2}
 \end{equation}

\begin{figure}
 \centering{\includegraphics[scale=0.275]{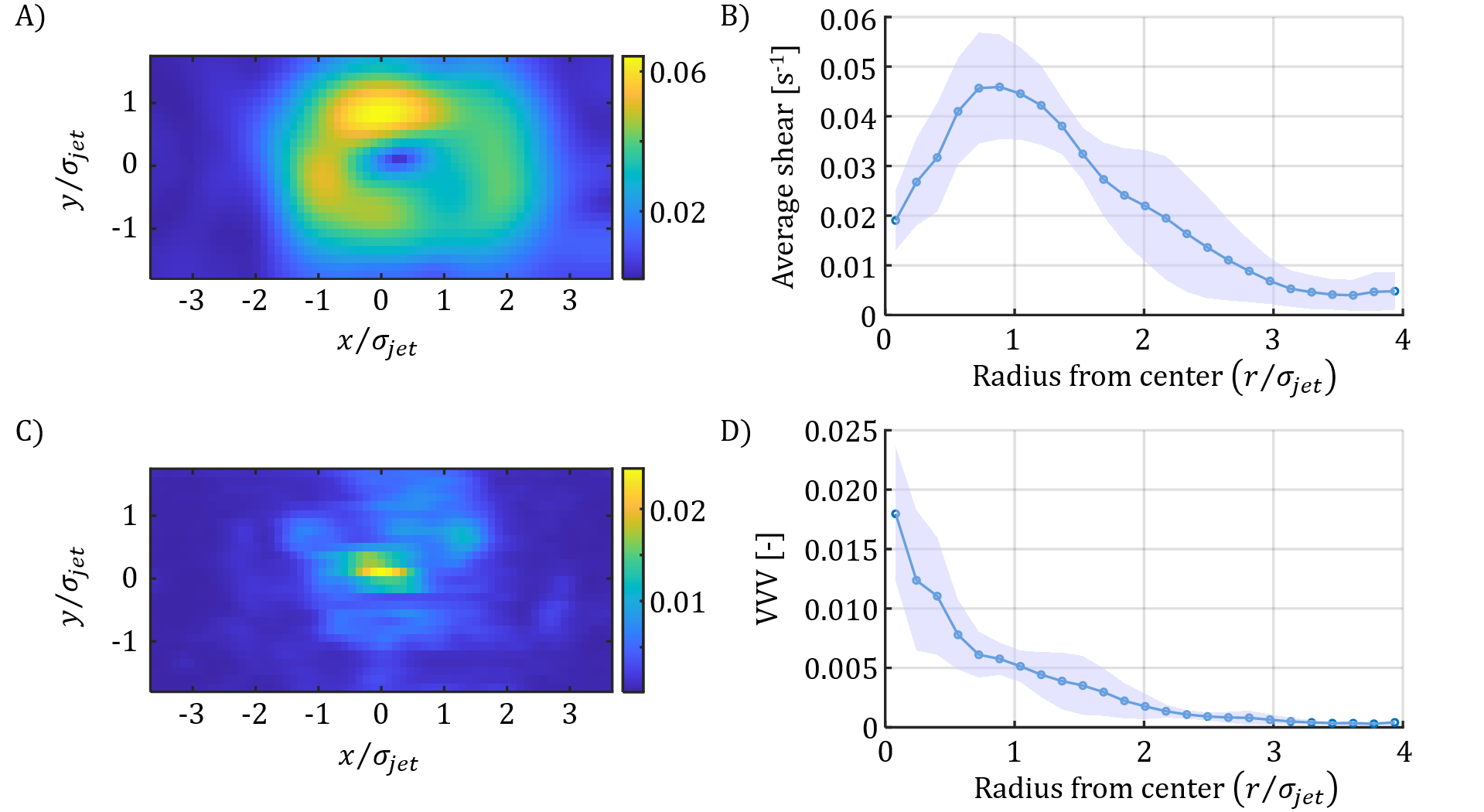}}
 \caption{Spatial characterization of shear and vertical velocity variance(VVV). A) Color map of shear magnitude, highlighting peak shear forming a distinct annular region. B) Radial distribution plot of shear magnitude versus radius, with shaded region representing mean ± standard deviation within each radial bin showing peak values approximately one characteristic jet radius from the center. C) Color map of vertically-averaged VVV distribution, illustrating the highest variability near the jet center. D) Radial distribution plot of vertically-averaged VVV versus radius, with shaded regions indicating mean ± standard deviation, showing rapid radial decay in unsteadiness.}
\label{fig:flow_metrics}
\end{figure}

This calculation was performed at each time step, and the results were subsequently time-averaged to produce a representative shear magnitude distribution (figure~\ref{fig:gaus_fit}A). Using the center coordinates of the 2-dimensional Gaussian fit as a reference point, the shear magnitudes were binned radially. Radial bins were defined as approximately one-quarter of the original horizontal grid bin resolution. The mean shear values and the corresponding standard deviations within each radial section were calculated to quantify the spatial variability and the uncertainty of the shear layer (figure~\ref{fig:flow_metrics}B).

The shear magnitude reached peak values of approximately 0.04–0.06 s$^{-1}$ at a radial distance that roughly corresponds to the characteristic radius of the jet, creating a distinct annular region. This spatial distribution of shear is closely aligned with typical jet behavior, where significant velocity gradients form near the jet boundary as a result of momentum exchange between the moving fluid core and the stationary surrounding fluid.

The vertical velocity variance (VVV) was derived from the fluctuations of the vertical velocity component around its mean. It is defined as: 

 \begin{equation}
     \text{VVV}= \frac{1}{2}\frac{w'^2}{w^2_{jet}},
 \end{equation}

where $w'$ denotes the variance of vertical velocity, or velocity fluctuations around the mean in time, and $w_{jet}$ is the characteristic jet centerline velocity previously defined. The variance of velocity fluctuations was evaluated over time, averaged vertically ($z$), and then radially binned similarly to the shear analysis (figure~\ref{fig:flow_metrics}).

The results revealed that VVV peaked at values between approximately 0.015 and 0.025 near the center of the jet, with a sharp decline to around 0.005 at distances corresponding to about one characteristic jet radius. 

\subsubsection{Swimmer flow interaction}

The vertical velocity field $w(x,y,z,t)$, planar shear $|\nabla w(x,y,t)|$, and vertical velocity variability VVV$(x,y)$ were interpolated onto the individual trajectories of the swimmers at each corresponding time step to explore the interactions between the swimmers and the induced flow structures. A swimmer-frame vertical swimming speed was calculated by subtracting the local flow velocity from the swimmer's vertical speed in the lab frame. 

Given the distinct radial symmetry of the flow structures, swimmer metrics were analyzed with respect to radial position. To reduce the potential bias of slower swimmers, whose longer residence times could skew the results, all metrics were calculated per swimmer. For radial analyses, the trajectories were classified based on characteristic jet radii, $\sigma_{jet}$: < 1$\sigma$, 1–2$\sigma$, 2–3$\sigma$, and > 3$\sigma$. The swimmer trajectories were segmented at these radial boundaries while preserving unique swimmer identifiers to avoid duplication in the swimmer count. Only trajectory segments exceeding 10 time steps (equivalent to 2 seconds) were included in the analysis. The resulting swimmer counts for each radial bin were as follows: 118 for < 1$\sigma$, 199 for 1–2$\sigma$, 71 for 2–3$\sigma$, and 21 for > 3$\sigma$. To statistically evaluate differences in swimmer metrics across radial bins, a one-way ANOVA was performed, with pairwise comparisons using the Tukey-Kramer multiple comparison test. Statistical significance was annotated on the basis of p-values: *** for $p<0.001$ and ** for $p<0.01$.

\begin{figure}
 \centering{\includegraphics[scale=0.275]{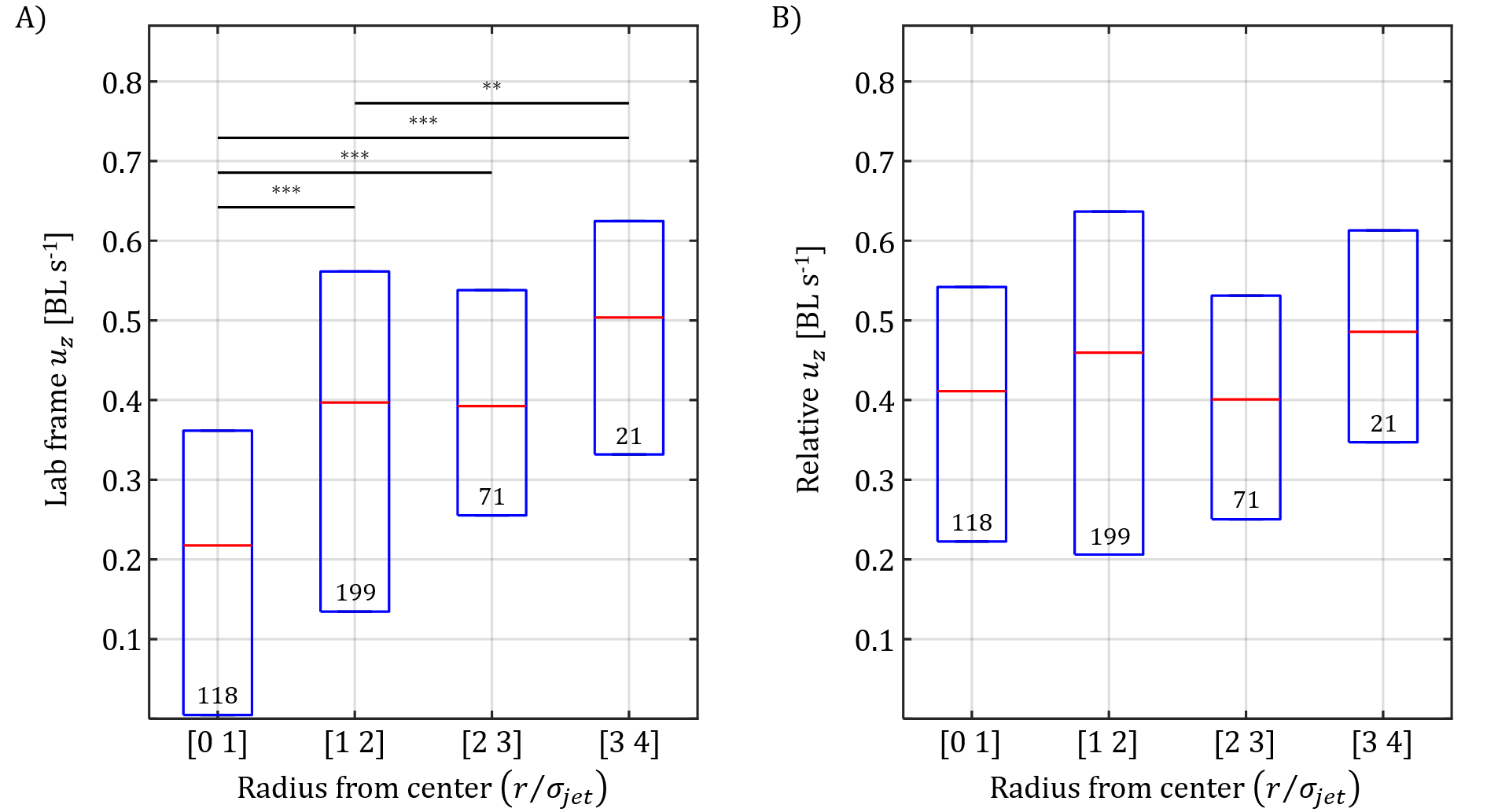}}
 \caption{Swimmer vertical speeds as a function of radial distance from the center of the tank. A) Box plots of laboratory-frame vertical swimming speeds (cm s$^{-1}$) across different radial bins. Boxes indicate the inter-quartile range, horizontal lines denote median values, and whiskers represent minimum and maximum values excluding outliers. B) Box plots of relative vertical swimming speeds, computed by subtracting the local vertical flow velocity from the swimmers' lab-frame speed. Relative swimmer speeds remained consistently between 0.4 and 0.5~cm~s$^{-1}$ across radial positions, demonstrating that the observed radial differences in panel A) were primarily due to local fluid velocities rather than active adjustments by swimmers to varying environmental conditions. Statistical significance between radial bins was assessed using one-way ANOVA and Tukey-Kramer multiple comparisons, with significant pairwise differences denoted by asterisks (** for p$<$0.01, *** for p$<$0.001). Sample sizes (n) for each radial bin are indicated at the bottom of each box plot.}
\label{fig:results_w_rel_box}
\end{figure}

An initial analysis of the vertical swimming speed in the laboratory frame suggested that the swimmers could exhibit an increasing vertical swimming speed with an increasing radial distance from the center of the tank (figure~\ref{fig:results_w_rel_box}A). However, when accounting for local flow velocity, the relative vertical swimming speeds remained consistently within the 0.4 to 0.5~cm~s$^{-1}$ range in all radial bins (Figure \ref{fig:results_w_rel_box}B). This suggests that swimmers do not actively adjust their speeds in response to reverse flow conditions or flow gradients.

\begin{figure}
 \centering{\includegraphics[scale=0.275]{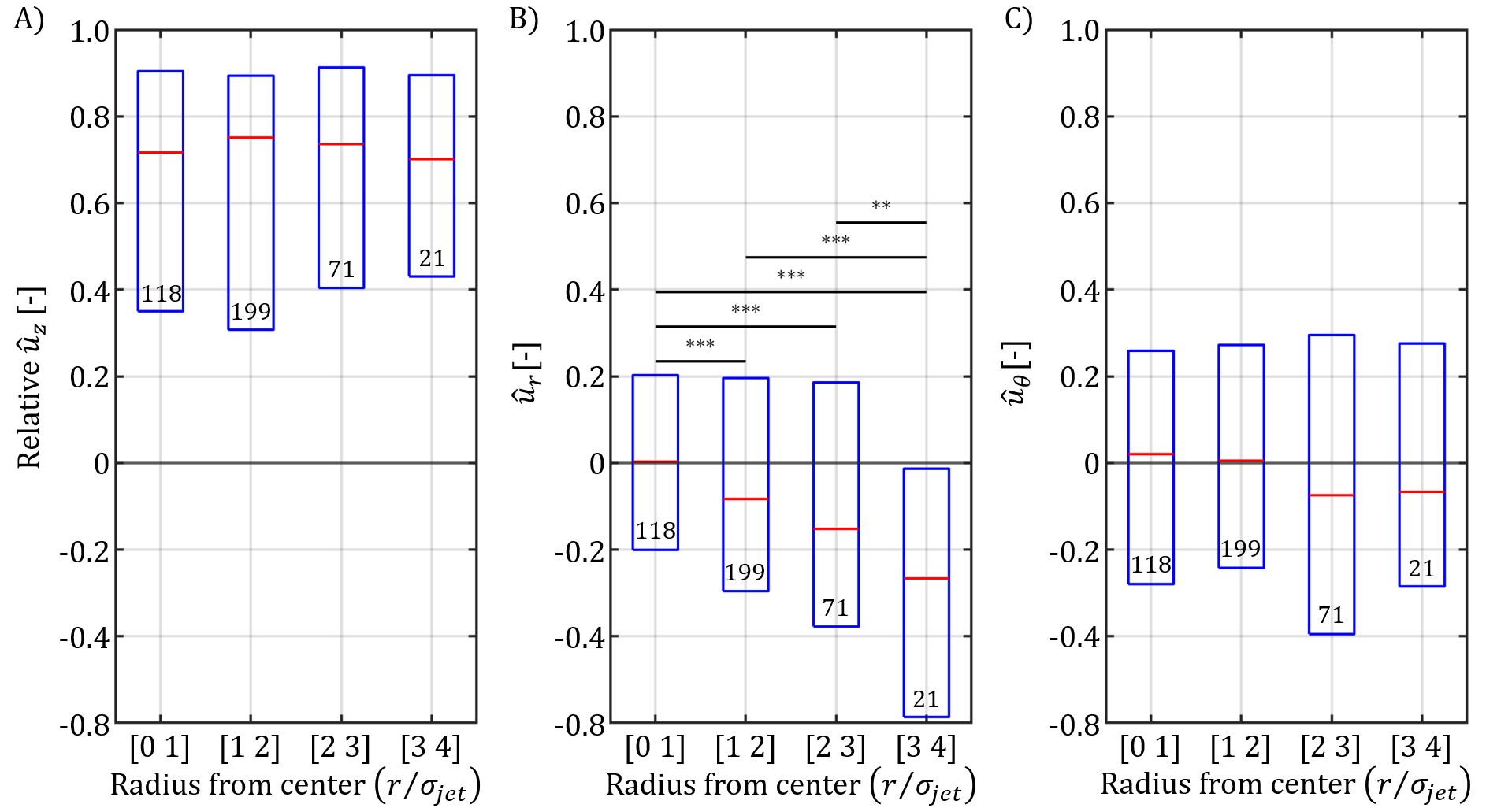}}
 \caption{Horizontal directional swimming behavior of brine shrimp. Box plots represent velocity unit vectors decomposed into vertical, radial, and azimuthal components across different radial bins from the tank center. A) Relative vertical unit vector components, indicating a consistently dominant vertical swimming. B) Azimuthal unit vector components, indicating no significant directional preference and consistent symmetrical or random swimming about the tank's central vertical axis. C) Radial unit vector components, showing increasingly negative (inward-directed) radial velocities with greater radial distances from the tank center. This inward-directed swimming behavior suggests positive phototaxis toward the centrally placed target flashlight, potentially in response to the radial illumination gradient. Statistical significance between radial bins was assessed using one-way ANOVA and Tukey-Kramer multiple comparisons, with significant pairwise differences denoted by asterisks (** for p$<$0.01, *** for p$<$0.001). Sample sizes (n) for each radial bin are indicated at the bottom of each box plot.}
\label{fig:results_unit_vec_box}
\end{figure}

Because of the dominant vertical swimming direction, there was no expected directional preference in the horizontal plane. This was evaluated by looking at the velocity unit vector in the radial and azimuthal components (figure~\ref{fig:results_unit_vec_box}). There was no significant directionality in azimuthal velocity, indicating symmetrical or random swimming about the tank's vertical axis. Radial velocity was 0 toward the center of the tank, indicating no directional preference; however, radial velocity became increasingly negative at larger distances from the center of the tank, indicating a preference for movement toward the tank center. This behavior likely stems from the positive phototaxis of brine shrimp, since they may sense the radial gradient in illumination created by the target flashlight. Interestingly, despite the inherent inward radial bias, the spatial distribution of swimmers has maintained a Gaussian-like profile. This stability in distribution may be a balance between inward-directed swimming driven by phototaxis and outward-directed dispersive behaviors, such as random reorientation, collisions, and interactions among swimmers. 

\begin{figure}
 \centering{\includegraphics[scale=0.275]{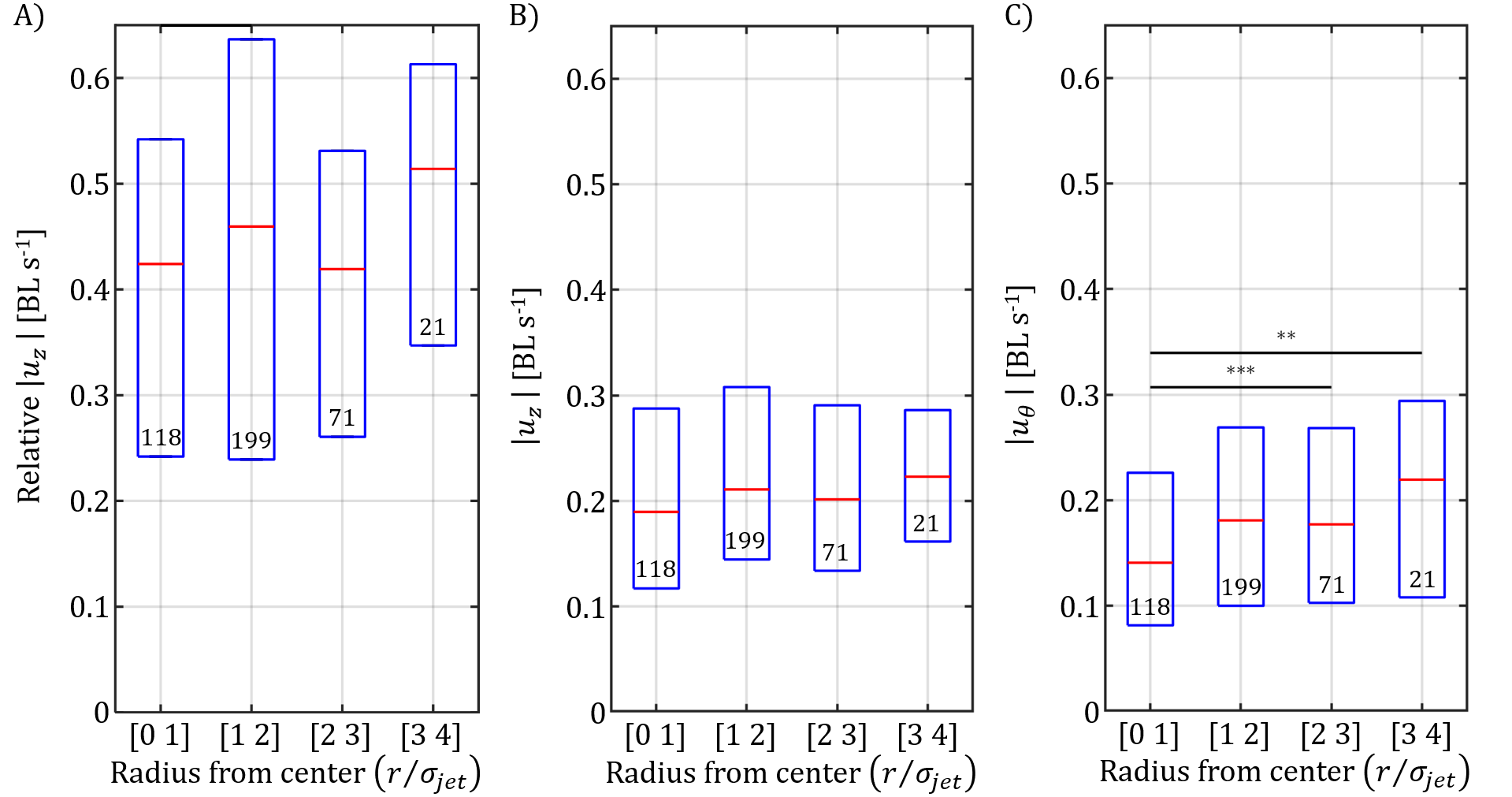}}
 \caption{Magnitude analysis of horizontal swimming velocity components of brine shrimp as a function of radial distance from the tank center. A) Vertical swimming speed showed no statistically significant differences across radial bins, indicating uniform vertical swimming throughout the observed region. B) Radial swimming speed showed no statistically significant differences across the radial bins, indicating uniform radial swimming activity throughout the observed region. C) Azimuthal swimming speed revealed a slight but noticeable decrease in swimmer velocity near the tank center, with a visual trend of increasing velocity magnitude at greater radial distances. Statistical significance between radial bins was assessed using one-way ANOVA and Tukey-Kramer multiple comparisons, with significant pairwise differences denoted by asterisks (** for p$<$0.01, *** for p$<$0.001). Sample sizes (n) for each radial bin are indicated below each box plot.}
\label{fig:results_mag_box}
\end{figure}

To complement this analysis of the azimuthal and radial velocity unit vectors, the magnitude, or speed,  of azimuthal and radial velocity in the horizontal plane is examined next. Vertical and radial speeds did not differ significantly between the various bins (figure~\ref{fig:results_mag_box}A,B). However, a slight decrease in azimuthal velocity near the center and a visual trend indicating an increase in azimuthal velocity with radial distance were observed (figure~\ref{fig:results_mag_box}C). 

To further elucidate these directional behaviors, the curvature of swimmer trajectories in the horizontal plane was computed as:

\begin{equation}
    \kappa_{xy} = \frac{|u_x a_y - u_y a_x|}{(u_x^2 + u_y^2)^{\frac{3}{2}}},
\end{equation}

where $u_x$,$u_y$ are horizontal swimming velocity components and $a_x$, $a_y$ are horizontal acceleration components. The radius of curvature, $R_{xy}$, was derived from the curvature:

\begin{equation}
    R_{xy}= \frac{1}{\kappa_{xy}}.
\end{equation}

\begin{figure}
 \centering{\includegraphics[scale=0.275]{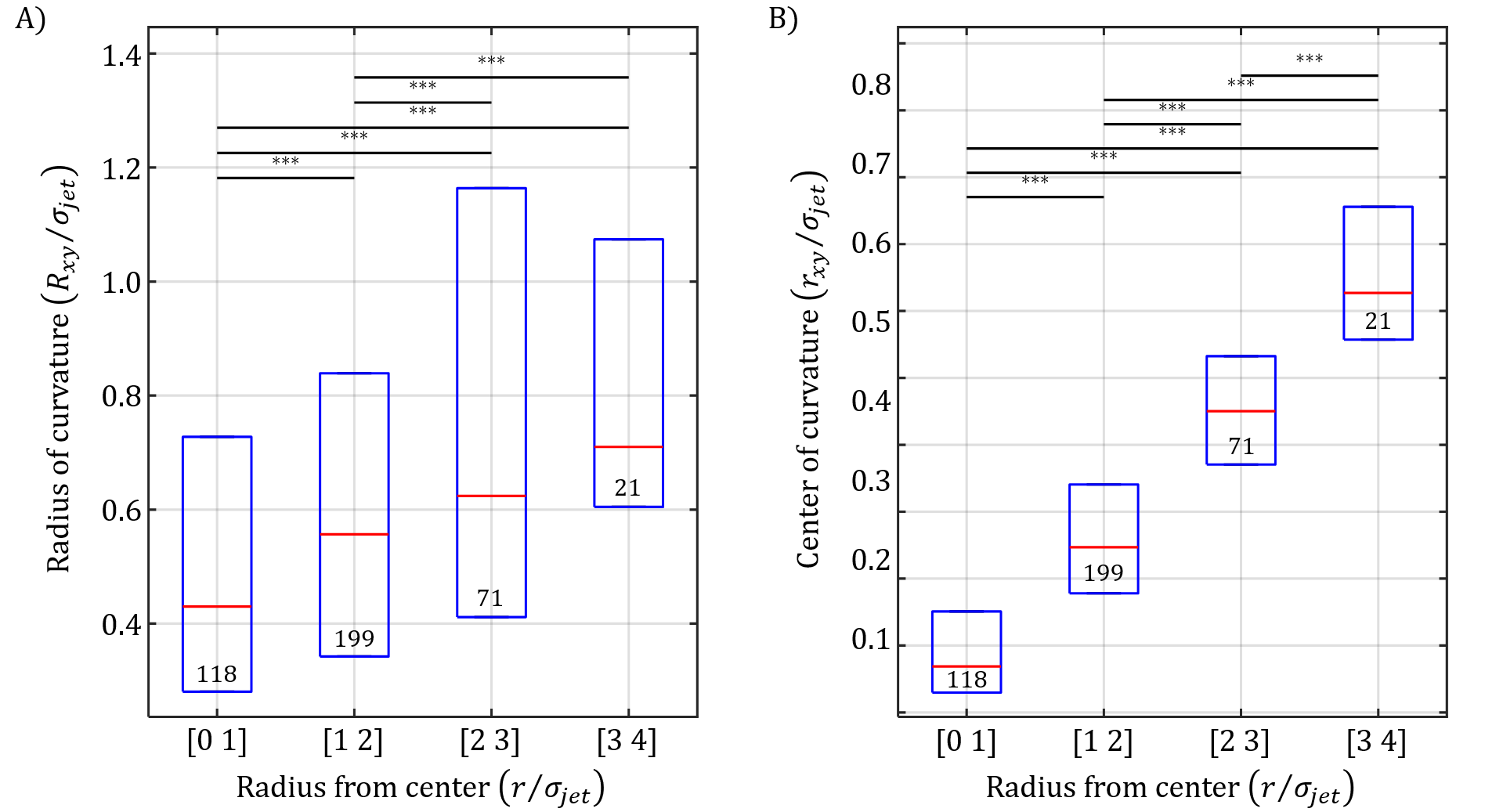}}
 \caption{Instantaneous curvature centers calculated from swimmer trajectories to identify potential collective vortical behavior. For each set of three sequential trajectory points, the perpendicular bisectors of the line segments connecting consecutive points were computed, with their intersection defining the instantaneous center of curvature. Collinear points were excluded from this analysis due to undefined curvature. A) Normalized radius of curvature, $R_{xy}/ \sigma_{jet}$, as a function of swimmer radial position, showing increasingly large circles formed by swimmers with increasing distance from the center. B) Radial position of instantaneous curvature centers, $r_{xy}/ \sigma_{jet}$ as a function of swimmer radial position, showing increasing deviation from the tank midpoint as swimmers move further from the center. Statistical significance between radial bins was assessed using one-way ANOVA and Tukey-Kramer multiple comparisons, with significant pairwise differences denoted by asterisks (** for p$<$0.01, *** for p$<$0.001). Sample sizes (n) for each radial bin are indicated below each box plot.
}
\label{fig:results_curv_box}
\end{figure}

The radius of curvature increased with radial distance, indicative of straighter trajectories further from the center. Coupled with increasing azimuthal velocities at larger radii, this suggests potential helical motion, characterized by larger, faster circles at outer radii coupled with consistent inward movement (figure~\ref{fig:results_curv_box}A).

To validate whether swimmers participated in organized vortical behavior~\citep{delcourt_collective_2016}, instantaneous curvature centers ($x_0$,$y_0$) were calculated by identifying intersections of perpendicular bisectors drawn from sets of three sequential trajectory points. Specifically, for three consecutive points $p_1$, $p_2$, $p_3$, the perpendicular bisectors were calculated from the midpoints of the segments connecting $p_1$ to $p_2$ and $p_2$ to $p_3$. The intersection of these bisectors provides the center of instantaneous curvature. Collinear points were excluded due to undefined curvature centers. The center of curvature was converted to a radial position $r_{xy}= \sqrt{x_0^2+y_0^2}$, such that if the swimmers collectively orbited a central vortex, the centers of curvature would cluster around the center of the tank ($r_{xy}=0$) across all radii. However, the curvature centers diverged significantly from the midpoint with increasing r moving outward radially (figure~\ref{fig:results_curv_box}B).

This analysis reveals that swimmers do not exhibit collective swirling or globally organized vortical motion. Instead, their trajectories represent locally organized but independent turning behaviors, which differ distinctly with radius. Closer to the center of the tank, swimmers demonstrate a higher curvature (frequent directional changes) at lower azimuthal speeds. At greater radial distances, swimmers move in smoother, less frequent turns at higher azimuthal speeds. These findings underscore autonomous local motion behaviors, ruling out large-scale coordinated vortical motion as the primary driver of observed swimmer trajectories.

Together, these analyses show that swimmers maintain a stable relative vertical swimming speed despite varying local flow conditions, indicating predominantly passive interaction with vertical flow structures. The inward radial preference, attributable to positive phototaxis, is balanced by dispersive behaviors (random reorientation, collisions, {$\sim$}1~cm exclusion zones), producing the observed Gaussian-like distribution. Horizontal trajectories show higher curvature and lower azimuthal speeds near the center, transitioning to smoother paths at larger radii; however, instantaneous curvature centers diverge from the tank midpoint with increasing radial distance, ruling out coordinated large-scale vortical motion. The close alignment of characteristic length scales for swimmer position ($\sigma \approx$ 2.9~cm), induced flow ($\sigma \approx$ 3.0~cm), and illumination (beam radius $\sim$3.5~cm) suggests that swimmers respond primarily to the illumination structure, with the induced flow inheriting that distribution.

\end{document}